\documentclass{article}

\usepackage{times}
\usepackage{graphicx} 
\usepackage{subfigure}

\usepackage{natbib}

\usepackage{algorithm}
\usepackage{algorithmic}

\usepackage[hyperfootnotes=false]{hyperref}

\usepackage[accepted]{arxiv}

\icmltitlerunning{Tuning the Molecular Weight Distribution from ATRP Using Deep RL}

\usepackage[version=3]{mhchem}
\usepackage{textcomp}
\usepackage{amsfonts}
\usepackage{stfloats}
\usepackage{bm}

\begin{document}

\twocolumn[
\icmltitle{Tuning the Molecular Weight Distribution from Atom Transfer Radical Polymerization Using Deep Reinforcement Learning}




\begin{icmlauthorlist}
\icmlauthor{Haichen Li}{cmu_chem,cmu_ml}
\icmlauthor{Christopher R. Collins}{cmu_chem}
\icmlauthor{Thomas G. Ribelli}{cmu_chem}
\icmlauthor{Krzysztof Matyjaszewski}{cmu_chem}
\icmlauthor{Geoffrey J. Gordon}{cmu_ml}
\icmlauthor{Tomasz Kowalewski}{cmu_chem}
\icmlauthor{David J. Yaron}{cmu_chem}
\end{icmlauthorlist}

\icmlaffiliation{cmu_chem}{Department of Chemistry, Carnegie Mellon University, Pittsburgh, PA 15213, USA}
\icmlaffiliation{cmu_ml}{School of Computer Science, Carnegie Mellon University, Pittsburgh, PA 15213, USA}

\icmlcorrespondingauthor{David J. Yaron}{yaron@cmu.edu}

\icmlkeywords{reinforcement learning, atom transfer radical polymerization, molecular distribution}

\vskip 0.3in
]



\printAffiliationsAndNotice{}  

\begin{abstract}
We devise a novel technique to control the shape of polymer molecular weight distributions (MWDs) in atom transfer radical polymerization (ATRP).  This technique makes use of recent advances in both simulation-based, model-free reinforcement learning (RL) and the numerical simulation of ATRP.  A simulation of ATRP is built that allows an RL controller to add chemical reagents throughout the course of the reaction.  The RL controller incorporates fully-connected and convolutional neural network architectures and bases its decision upon the current status of the ATRP reaction.  The initial, untrained, controller leads to ending MWDs with large variability, allowing the RL algorithm to explore a large search space.   When trained using an actor-critic algorithm, the RL controller is able to discover and optimize control policies that lead to a variety of target MWDs.  The target MWDs include Gaussians of various width, and more diverse shapes such as bimodal distributions.  The learned control policies are robust and transfer to similar but not identical ATRP reaction settings, even under the presence of simulated noise.  We believe this work is a proof-of-concept for employing modern artificial intelligence techniques in the synthesis of new functional polymer materials.
\end{abstract}

\newcommand{\fig}{Figure}


\section{Introduction}
Most current approaches to development of new materials follow a sequential, iterative process that requires extensive human labor to synthesize new materials and elucidate their properties and functions.  Over the next decades, it seems likely that this inherently slow and labor intensive approach to chemical research will be transformed through the incorporation of new technologies originating from computer science, robotics, and advanced manufacturing.\cite{dragan2017learning,boots2011closing}  A central challenge is finding ways to use these powerful new technologies to guide chemical processes to desired outcomes.\cite{ley2015organic}  Recent advances in reinforcement learning (RL) have enabled computing systems to guide vehicles through complex simulation environments,\cite{koutnik2014evolving} and select moves that guide games such as Go and chess to winning conclusions.\cite{mnih2015human,silver2016mastering,silver2017mastering,silver2017mastering2}  For chemical problems, RL has been used to generate candidate drug molecules in a \textit{de novo} manner,\cite{popova2017deep,olivecrona2017molecular} and to optimize reaction conditions for organic synthesis.\cite{zhou2017optimizing}  This work investigates the benefits and challenges of using RL to guide chemical reactions towards specific synthetic targets.  The investigation is done through computational experiments that use RL to control a simulated reaction system, where the simulation models the chemical kinetics present in the system.

In this work, the simulated reaction system is that of atom transfer radical polymerization (ATRP).\cite{matyjaszewski2012atom,matyjaszewski2001atom,matyjaszewski2014macromolecular,hawker1994molecular}  ATRP is among the mostly widely used and effective means to control the polymerization of a wide variety of vinyl monomers.  ATRP allows the synthesis of polymers with predetermined molecular weights, narrow molecular weight distributions (MWDs),\cite{di2010transition} and adjustable polydispersity.\cite{plichta2012tuning,lynd2005influence,lynd2007effects,lynd2007role,lynd2008theory,listak2008effect,gentekos2016beyond}  The high degree of control allows the synthesis of various polymeric architectures \cite{matyjaszewski2005controlled} such as block copolymers,\cite{min2005preparation,carlmark2003atrp,majewski2015millisecond,majewski2015arbitrary} star polymers,\cite{miura2005synthesis,gao2006synthesis,li2004multicompartment} and molecular brushes.\cite{gao2007synthesis}  Temporal and spatial control has also been applied in ATRP to further increase the level of control over the polymerization.\cite{wang2017enhancing,wang2017temporal,ribelli2014radicals,dadashi2017photoinduced}  More recently, chemists have been working on ways to achieve MWDs with more flexible forms,\cite{gentekos2016beyond,carmean2017ultra} as this may provide a means to tailor mechanical and processability of the resulting plastics.\cite{kottisch2016shaping}

\newcommand{\CuI}{\ce{L/Cu^{I}}}
\newcommand{\CuII}{\ce{L/Cu^{II}-Br}}
\begin{figure}[ht]
  \centering
  \includegraphics[width=0.9\columnwidth]{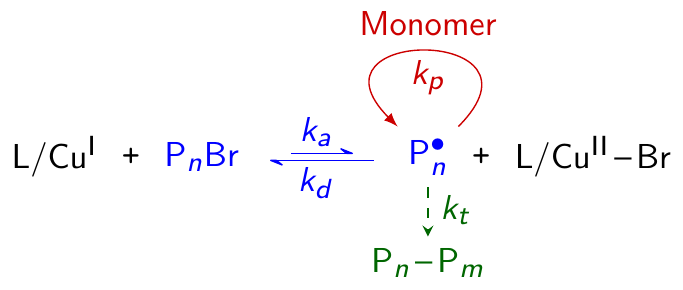}
  \caption{Reaction mechanism of ATRP. Polymer species include radical chains \ce{P^$\bullet$_{n}} and dormant chains \ce{P_nBr} with reduced chain length n and chains that terminated through recombination \ce{P_n-P_m}. \CuI~and~\CuII~are ATRP catalysts, where L represents the ligand.  $k_p$, $k_a$, $k_d$, and $k_t$ are kinetic rate constants for chain propagation, activation, deactivation, and termination, respectively.}
  \label{fig:atrp_scheme}
\end{figure}

\begin{figure}[ht]
  \centering
  \includegraphics[width=0.9\columnwidth]{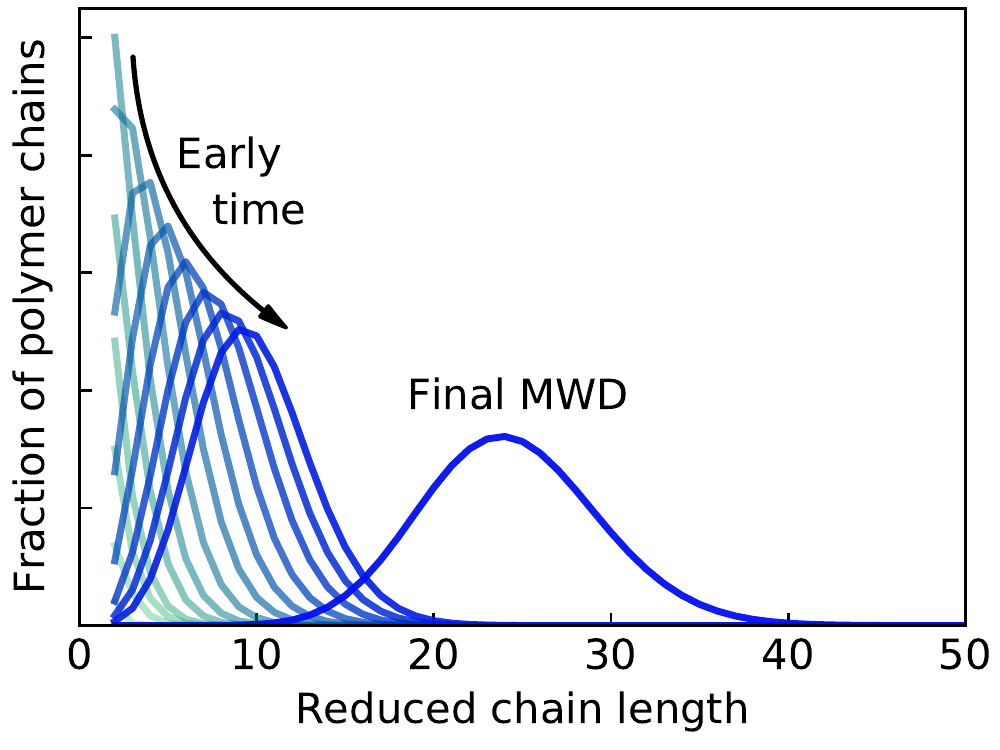}
  \caption{Evolution of polymer MWD in a simulated ATRP reaction.}
  \label{fig:evolution}
\end{figure}

In addition to its importance, ATRP is well suited to the computational experiments carried out here.  The chemical kinetics of ATRP are shown schematically in \fig~\ref{fig:atrp_scheme}. Control of the polymerization process is related to the activation, $k_a$, and deactivation, $k_d$, reactions which inter-convert dormant chains, \ce{P_nBr}, and active, free radical chains, \ce{P^$\bullet$_{n}}. The active chains grow in length through propagation reactions, $k_p$. The equilibrium between dormant and active chains can be used to maintain a low concentration of active chains, leading to more controlled growth and a reduction in termination reactions, $k_t$, that broaden the final MWD. These kinetics are sufficiently well understood\cite{goto2004kinetics,tang2006effect} that simulations provide reliable results.\cite{weiss2015atom,preturlan2016numerical,drache2012simulating,vieira2016optimization,van2012linear,d2012kinetic,krys2017mechanism,krys2017kinetics,zhong2013reversible} It is also computationally feasible to carry out a large number of simulated reactions.  \fig~\ref{fig:evolution} shows how the MWD evolves in a single reaction simulation, which finishes in about 1 minute on a 2.4\,GHz CPU core.  MWDs will be shown as the fraction of polymer chains (vertical axis) with a specific reduced chain length (horizontal axis), where the reduced chain length refers to the number of monomers incorporated into the chain.

ATRP reactions can also be manipulated in a large variety of ways because of the multiple interacting chemical reactions, and the shape of the MWD provides a diverse set of targets.  This makes the system a good choice for evaluating the degree to which RL can guide a chemical process to a desired synthetic target.  ATRP reactions are typically carried out by creating an initial mixture of chemical reagents and keeping the temperature and other reaction conditions steady.  However, a greater diversity of MWDs can be obtained by taking actions, such as adding chemical reagents, throughout the polymerization process.\cite{gentekos2016beyond}  Here, we use RL to decide which actions to take, based on the current state of the reaction system.  In this manner, it is analogous to having a human continuously monitor the reaction and take actions that guide the system towards the target MWD.  This use of a state-dependent decision process is a potential advantage of using RL.  Consider an alternative approach in which the simulation is used to develop a protocol that specifies the times at which to perform various actions.  Such a protocol is likely to be quite sensitive to the specific kinetic parameters used in the simulation.  The RL controller may lower this sensitivity by basing its decisions on the current state of the reaction system.  Below, the current state upon which the RL controller makes its decisions includes the current MWD.  The controller is then expected to succeed provided the correct action to take at a given time depends primarily on the difference between the current MWD and the target MWD (\fig~\ref{fig:evolution}), as opposed to the specific kinetic parameters.  Ideally, an RL algorithm trained on a simulated reaction may be able to succeed in the real laboratory with limited additional training, provided the simulated reaction behaves like the actual one.  Such transfer from simulated to real-world reactions is especially important given the potentially large number of reaction trials needed for training, and the inherent cost of carrying out chemical experiments.  In our computational experiments, we assess the sensitivity to the simulation parameters by including noise in both the kinetic parameters used in the simulation and in the states of the current reaction system.

\begin{figure}[ht]
  \centering
  \includegraphics[width=0.95\columnwidth]{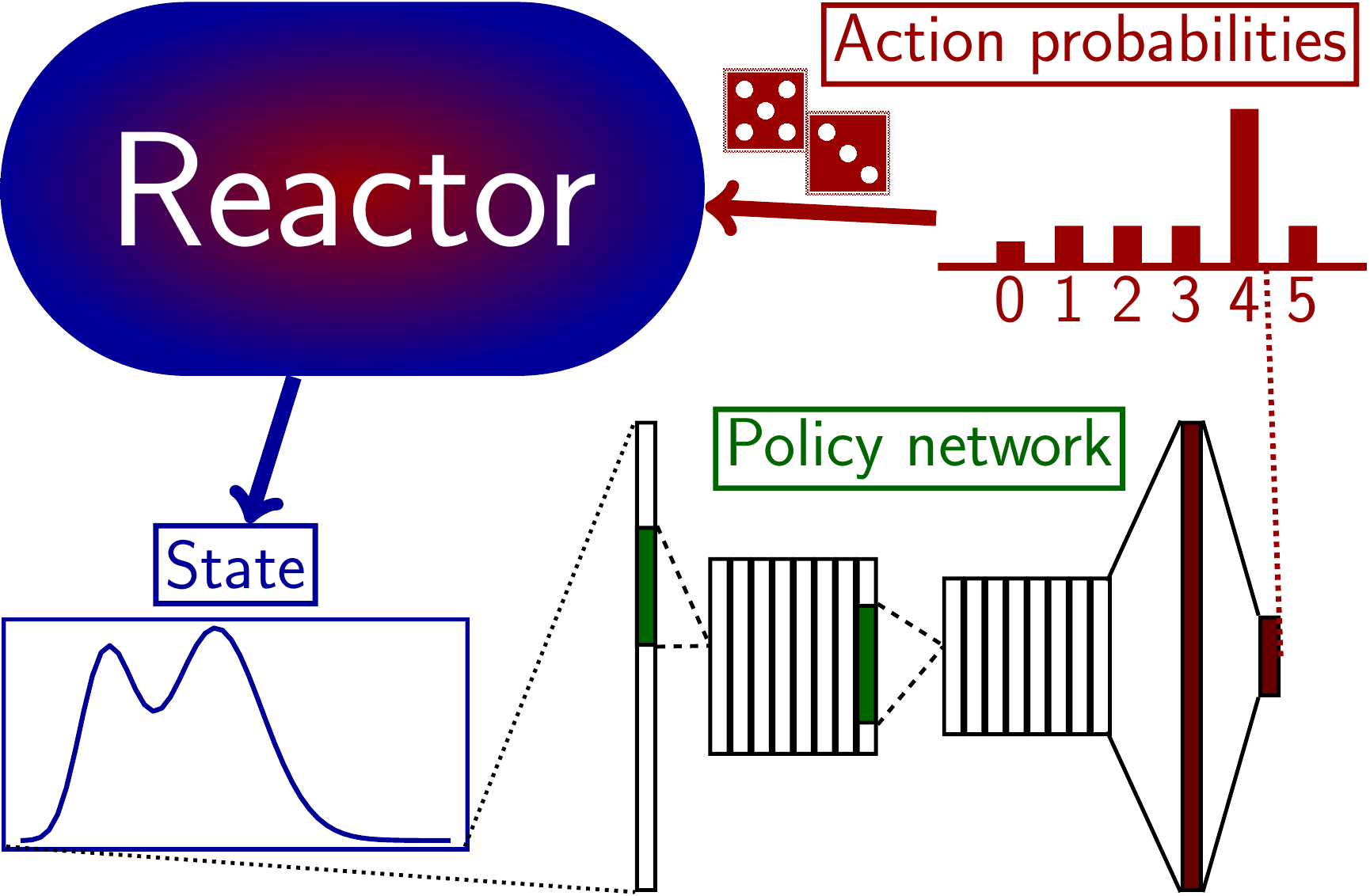}
  \caption{Flow chart showing how the policy network of the RL controller selects actions to apply to the simulated ATRP reactor.}
  \label{fig:control_flow_chart}
\end{figure}

\fig~\ref{fig:control_flow_chart} provides a schematic view of the RL controller.  The current state is fed into the RL controller (policy network), which produces a probability distribution for each of the available actions.  An action is then drawn from this probability distribution, and performed on the reactor.  The design of the RL controller is inspired by recent advances in deep reinforcement learning,\cite{li2017deep,arulkumaran2017brief,henderson2017deep} which use neural networks for the policy network and other components.  The combination of modern deep learning models, represented by convolutional neural networks,\cite{krizhevsky2012imagenet,deng2013new,lecun2015deep,schmidhuber2015deep} and efficient RL algorithms\cite{gordon2001reinforcement,gordon1995stable} such as deep Q-learning,\cite{mnih2015human,van2016deep,liang2016state} proximal policy methods,\cite{schulman2017proximal} and asynchronous advantage actor-critic (A3C)\cite{mnih2016asynchronous,fortunato2017noisy} has lead to numerous successful applications in control tasks with large state spaces.\cite{lillicrap2015continuous,roy2003exponential,dragan2017learning}  The computational experiments presented here examine the use of modern deep reinforcement learning techniques to guide chemical synthesis of new materials.

\section{Related works}
There have been many studies that control the state and dynamics of chemical reactors based on classical control theory.\cite{nomikos1994monitoring}  Model-based controllers,\cite{binder2001introduction} some of which employ neural networks,\cite{hussain1999review} have been developed for a number of control tasks involving continuous stirred tank reactors,\cite{ydstie1990forecasting,lightbody1995direct,yang1994adaptive,watanabe1994comparison,bahita2016model,galluzzo2011control} batch processes,\cite{srinivasan2003dynamica,srinivasan2003dynamicb,nie2012integrated} hydrolyzers,\cite{lim2010control} bioreactors,\cite{bovskovic1995comparison,chovan1996neural,de2016soft} pH neutralization processes,\cite{nahas1992nonlinear,mahmoodi2009nonlinear,hermansson2015model,nejati2012comparison} strip thickness in steel-rolling mills,\cite{sbarbaro1993neural} and system pressure.\cite{turner1995neural}  Model-free controllers trained through RL also exist for controlling chemical processes such as neutralization\cite{syafiie2007model} and wastewater treatment\cite{syafiie2011model} or chemical reactor valves.\cite{de2012novel}

Due to its industrial importance, polymer synthesis has been a primary target for the development of chemical engineering controllers.\cite{chatzidoukas2003optimal}  Some of these make use of neural networks to control the reactor temperature in the free radical polymerization of styrene.\cite{hosen2011control}  McAfee et al.  developed an automatic polymer molecular weight controller\cite{mcafee2016automatic} for free radical polymerization.  This controller is based on online molar mass monitoring techniques\cite{florenzano1998absolute} and is able to follow a specific chain growth trajectory with respect to time by controlling the monomer flow rate in a continuous flow reactor.  Similar online monitoring techniques have recently enabled controlling the modality of free radical polymerization products,\cite{leonardi2017automatic} providing optimal feedback control to acrylamide-water-potassium persulfate polymerization reactors,\cite{ghadipasha2017online} and monitoring multiple ionic strengths during the synthesis of copolymeric polyelectrolytes.\cite{wu2017simultaneous}  However, none of these works attempted to control the precise shape of polymer MWD shapes, nor did they use an artificial intelligence (AI) driven approach to design new materials.  The significance of this work lies in that it is a first trial of building an AI agent that is trained \textit{tabula rasa} to discover and optimize synthetic routes for human-specified, arbitrary polymer products with specific MWD shapes.  Another novel aspect of the current work is the use of a simulation to train a highly-flexible controller, although the transfer of this controller to actual reaction processes, possibly achievable with modern transfer learning\cite{taylor2009transfer,pan2010survey,wang2014flexible,christiano2016transfer,barrett2010transfer} and imitation learning techniques,\cite{ross2011reduction,sun2017deeply} is left to future work.

\section{Methodology}
\subsection{Simulating ATRP}

\newcommand{\Mono}{\ce{M}}
\newcommand{\Rad}[1]{\ce{P^\bullet_{#1}}}
\newcommand{\Dorm}[1]{\ce{P_{#1}Br}}
\newcommand{\Ter}[1]{\ce{T_{#1}}}
\newcommand{\cMono}{[\Mono]}
\newcommand{\cCuI}{[\ce{Cu^{I}}]}
\newcommand{\cCuII}{[\ce{Cu^{II}}]}
\newcommand{\cRad}[1]{[\Rad{#1}]}
\newcommand{\cDorm}[1]{[\Dorm{#1}]}
\newcommand{\cTer}[1]{[\Ter{#1}]}
\newcommand{\sumRad}{\sum_{i=1}^N{\cRad{i}}}
\newcommand{\sumDorm}{\sum_{i=1}^N{\cDorm{i}}}

\renewcommand{\arraystretch}{1.6}
\begin{table*}[bp]
    \centering
    \caption{ATRP kinetics equations.  \ce{Cu^{I}} and \ce{Cu^{II}} stand for the ATRP activator and deactivator \CuI\,and \CuII, respectively.}
    \label{tab:atrp_kinetics}
    \begin{tabular}{cl}
    \hline
    Monomer & \(\cMono' = -k_p\cMono\sumRad \) \\
    Activator & \(\cCuI' = k_d\cCuII\sumRad - k_a\cCuI\sumDorm \) \\
    Deactivator & \(\cCuII' = k_a\cCuI\sumDorm - k_d\cCuII\sumRad \) \\
    Dormant chains & \(\cDorm{n}' = k_d\cCuII\cRad{n} - k_a\cCuI\cDorm{n},~1 \leq n \leq N \) \\
    Smallest radical & \(\cRad{1}' = -k_p\cMono\cRad{1} + k_a\cCuI\cDorm{1} -k_d\cCuII\cRad{1} - 2k_t\cRad{1}\sumRad \) \\
    Other radicals & \(\cRad{n}' = k_p\cMono(\cRad{n-1} - \cRad{n}) + k_a\cCuI\cDorm{n} -k_d\cCuII\cRad{n} - 2k_t\cRad{n}\sumRad,~2 \leq n \leq N\) \\
    Terminated chains & \(\cTer{n}' = \sum_{i=1}^{n-1}k_t\cRad{i}\cRad{n-i},~2 \leq n \leq 2N \) \\
    \hline
    \end{tabular}
\end{table*}

We select styrene ATRP as our simulation system.  Simulation of styrene ATRP may be done by solving the ATRP chemical kinetics ordinary differential equations (ODEs) in Table~\ref{tab:atrp_kinetics},\cite{weiss2015atom,preturlan2016numerical,vieira2016simulation,li2011kinetics} by method of moments,\cite{zhu1999modeling} or by Monte Carlo methods.\cite{al2006dynamic,najafi2010comprehensive,najafi2011simulation,turgman2012computer,payne2013arget,toloza2013theoretical}  This work directly solves the ODEs because this allows accurate tracking of the concentration of individual polymer chains while being more computationally efficient than Monte Carlo methods.

In the ODEs of Table~\ref{tab:atrp_kinetics}, $\Mono$ is monomer; $\Rad{n}$, $\Dorm{n}$, and $\Ter{n}$ represent {length-$n$} radical chain, dormant chain, and terminated chain, respectively.  $\Dorm{1}$ is also the initiator of radical polymerization.  $k_p$, $k_a$, $k_d$, and $k_t$ are propagation, activation, deactivation, and termination rate constants, respectively.  $N$ is the maximum allowed dormant/radical chain length in the numerical simulation.  Consequently, the maximum allowed terminated chain length is $2N$, assuming styrene radicals terminate via combination.\cite{nakamura2016mechanism}  We set $N=100$ in all ATRP simulations in this work.  This number is sufficiently large for our purpose as the lengths of dormant or terminated chains do not exceed 75 or 150, respectively, in any of the simulations.  We used a set of well-established rate constants based on experimental results of the ATRP of bulk styrene at 110~\textdegree{C} (383.15 K) using dNbpy as the ligand\cite{wang1995controlled,matyjaszewski2001atom,patten1996polymers,matyjaszewski1997controlled}: $k_p = 1.6 \times 10^3$, $k_a = 0.45$, $k_d = 1.1 \times 10^7$, and $k_t = 10^8$ (units are $\text{M}^{-1}\text{s}^{-1}$).  It was assumed the reactor remained at this temperature for the duration of the polymerization.  Although the rate constants depend on the degree of polymerization,\cite{gridnev1996dependence} we assumed the same rate constants for polymer chains with different lengths.  This assumption does not bias the nature of ATRP qualitatively and has been practiced in almost all previous ATRP simulation research.\cite{weiss2015atom,preturlan2016numerical,vieira2016simulation,li2011kinetics,vieira2015styrene}  In some of our simulations, we altered the rate constants by up to $\pm 30\%$ to account for possible inaccuracies in the measurement of these values and other unpredictable situations such as turbulence in the reactor temperature.  We employed the VODE\cite{brown1989vode,byrne1975polyalgorithm,hindmarsh1977episode,hindmarsh1983odepack,jackson1980alternative} integrator implemented in SciPy~0.19 using a maximum internal integration step of 5000, which is sufficient to achieve final MWDs with high accuracy.  We chose the ``backward differentiation formulas'' integration method because the ODEs are stiff.

In practice, styrene ATRP is close to an ideal living polymerization,\cite{patten1996polymers,matyjaszewski1997controlled} with termination playing only a small role in establishing the final MWD.  Excluding termination from the simulation reduces the the total number of ODEs by about $2/3$ and substantially reduces the computer time needed for the simulation.  Therefore, in most of the cases, we train the RL agents on \emph{no-termination} environments to save computational cost.  Note that we still evaluate their performance on \emph{with-termination} environments.  Moreover, this strategy allows us to test the transferability of control policies learned by the RL agent onto similar but not identical environments, which could be of great importance in later works where we need to apply control policies learned with simulated environments to real, physically built reactors.

We assume that the volume of the system is completely determined by the amount of solvent and the number of monomer equivalents (including monomers incorporated in polymer chains).  To calculate the system volume, we use a bulk styrene density of 8.73 mol/L as reported in early works\cite{weiss2015atom} and a solvent density of 1.00 mol/L.

\subsection{Using RL to control the ATRP reactor simulation}
A reinforcement learning problem is usually phrased as an agent interacting with an environment (\fig~\ref{fig:rl_diagram}).  In our case, the agent is an RL controller and the environment is the ATRP reactor simulator.  The agent interacts with the simulation at times separated by constant intervals, $t_\textit{step}$.  The interaction between the agent and the environment consists of three elements, each of which is indexed by the timestep (shown as a subscript $t$):
\begin{figure}[ht]
  \centering
  \includegraphics[width=0.95\columnwidth]{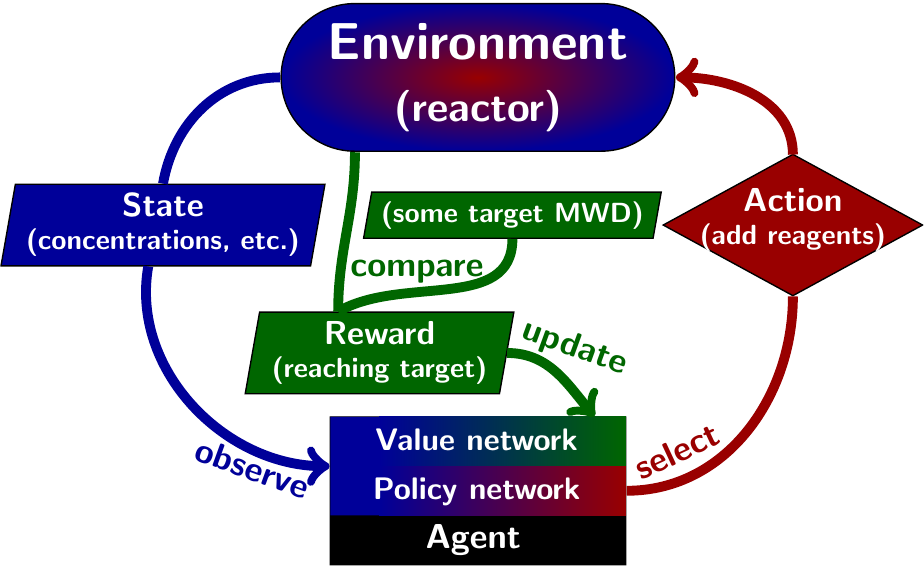}
  \caption{A schematic diagram of applying deep reinforcement learning in the ATRP reactor control setting.}
  \label{fig:rl_diagram}
\end{figure}
\begin{description}
\item[State ($\bm{s_t}$)]  At each timestep, the agent is given a vector, $s_t$, that is interpreted as the current state of the reaction system.  The state vector is used by the agent to select actions.  Here, $s_t$ includes: (i) the concentrations of the non-trace species: monomer, dormant chains (${\Dorm{1},\,\cdots,\,\Dorm{N}}$), and \ce{Cu}-based ATRP catalysts, (ii) the volume of the solution, and (iii) binary indicators of whether each of the addable reagents has reached its budget.  Note that we include the monomer quantity into the state vector by adding it onto the quantity of the initiator, or the shortest dormant chain.
\item[Action ($\bm{a_t}$)]  The agent is given a set of actions, $\mathcal{A}$, from which to select an action, $a_t$, to apply at timestep $t$.  The set of actions is fixed and does not change throughout the simulation.  Here, the actions correspond to the addition of a fixed amount of a chemical reagent.  The set of actions, $\mathcal{A}$, also includes a \emph{no-op}, selection of which means that no action is taken on the reaction simulation environment.  The addable reagents are listed in Table~\ref{tab:amounts}, along with the amount that is added when the action is selected and the budget.  When a reagent reaches its budget, the agent may still select the corresponding action, but this action becomes a no-op and does not alter the reaction simulation environment.  Although the simulation allows addition of solvent, the effects of this action are not examined here.  A very small amount of solvent is, however, used to initialize the simulation with a non-zero volume of a non-reactive species.  Inclusion of other actions, such as changes in temperature, are possible but these are also not examined here.
\item[Reward ($\bm{r_t}$)]  At each timestep, the agent is given a reward, $r_t$, that indicates the degree to which the agent is succeeding at its task.  In many RL problems, rewards may accrue at any time point.  Here, however, the reward is based on the final MWD and so the agent receives a reward only when the reaction has run to completion.  In practice, we allow the agent to interact with the simulation until all addable reagents have reached their budgets.  The simulation then continues for a terminal simulation time of $t_\textit{terminal} = 10^5$ seconds.  The simulation environment then provides a reward to the agent based on the difference between the ending dormant chain MWD and the target MWD.  This reward is defined in a two-level manner: when the maximum absolute difference between the normalized ending MWD and target MWD is less than $1\times10^{-2}$ the agent obtains a reward of 0.1, and when this difference is less than $3\times10^{-3}$, the agent obtains a reward of 1.0.  This two-level reward structure was determined empirically, with the lower first-level reward helping guide the agent in the early stages of training.
\end{description}

\begin{table}[ht!]
\small
\centering
\caption{The initial amounts, addition unit amounts, and budget limits used for simulating styrene ATRP in this work.  All quantities are in units of \textit{mol}.}
\begin{tabular}{cccc}
\hline\hline
Addable reagents    & Initial   & Addition unit & Budget limit \\
\hline
Monomer     & 0         & 0.1           & 10.0  \\
Activator   & 0         & 0.004         & 0.2   \\
Deactivator & 0         & 0.004         & 0.2   \\
Initiator   & 0         & 0.008         & 0.4   \\
Solvent     & 0.01      & 0             & 0     \\
\hline
\end{tabular}
\label{tab:amounts}
\end{table}

A single simulated ATRP reaction corresponds, in RL, to a single \emph{episode}.  Each episode begins with a small amount of solvent (Table~\ref{tab:amounts}) and iterates through steps in which the agent is given the current state, $s_t$, the agent selects an action $a_t$ that is applied to the simulation, and the simulation then runs for a time $t_\textit{step}$.  When all addable reagents have reached their budgets, the simulation continues for $t_\textit{terminal} = 10^5$ seconds and returns a reward based on the difference between the ending dormant chain MWD and the target MWD.

\newcommand{\pol}[1]{\pi_{\theta_p}(s_{#1})}
\newcommand{\locpol}[1]{\pi_{\theta_p'}(s_{#1})}
\newcommand{\pola}[1]{\pi_{\theta_p}(a_{#1}|s_{#1})}
\newcommand{\locpola}[1]{\pi_{\theta_p'}(a_{#1}|s_{#1})}
\newcommand{\approxval}[1]{V_{\theta_v}(s_{#1})}
\newcommand{\trueval}{\mathcal{V}(s_t)}
\newcommand{\approxvalloc}[1]{V_{\theta_v'}(s_{#1})}
\newcommand{\entropy}{H\big(\locpol{t'}\big)}
\newcommand{\indic}{I_{t+L}}

To train the agent, we use the A3C algorithm, a recent advance in actor-critic methods\cite{degris2012model} that achieved state-of-the-art performance on many discrete-action control tasks.\cite{rusu2016sim}  Actor-critic\cite{konda2000actor} algorithms are a subclass of RL algorithms based on simultaneous training of two functions:
\begin{description}
\item[Policy ($\bm{\pi}_{\bm{\theta_p}}\bm{(s_t)}$)]  The policy is used to select actions, e.g., which chemical reagent to add at time $t$.  As shown schematically in \fig~\ref{fig:control_flow_chart}, actions are drawn from a probability distribution.  The policy function generates this probability distribution, $\pola{t}$, which specifies, given the state of the ATRP reactor $s_t$, the probability that action $a_t$ should be selected. The subscript $\theta_p$ represents the set of parameters that parameterize the policy function.  In A3C, where a neural network is used for the policy, $\theta_p$ represents the parameters in this neural network.\cite{sutton2000policy,greensmith2004variance}
\item[Value ($\bm{V}_{\bm{\theta_v}}\bm{(s_t)}$)]  Although the policy function is sufficient for use of the RL controller, training also involves a value function, $\approxval{t}$.  Qualitatively, this function is a measure of whether the reaction is on track to generate rewards.  More precisely, we define a \emph{return} ${R_t=\sum_{t'=t}^{T}\gamma^{t'-t}r_{t'}}$ which includes not only the reward at the current state, but also future states up to timestep $T$.  This is especially relevant here, as rewards are based on the final MWD and so are given only at the end of a reaction.  A factor $\gamma$, which is greater than 0 and less than 1, discounts the reward for each step into the future, and is included to guarantee convergence of RL algorithms.  The value function, $\approxval{t}$, approximates the expected return, $\mathbb{E}[R_t|s_t]$, from state $s_t$.  A3C uses a neural network for the value function, and $\theta_v$ represents the parameters in this network.
\end{description}
Below, we compare results from two different neural network architectures, labeled FCNN and 1D-CNN (see Section~\ref{ssec:imple}).

During training, A3C updates the parameters, $\theta_p$ and $\theta_v$, of the policy and value functions. The actor-critic aspect of A3C refers to the use of the value function to critique the policy's ability to select valuable actions. To update $\theta_p$, policy gradient steps are taken according to the direction given by ${\nabla_{\theta_p}\log\pola{t}\big(R_{t} - \approxval{t}\big)}$.  Note that the current value function, $\approxval{t}$, is used to update the policy, with the policy gradient step being in a direction that will cause the policy to favor actions that maximize the expected return.  This may be viewed as using the value function to critique actions being selected by the policy.  Moreover, the policy gradient becomes more reliable when the value function estimates the expected return more accurately.  To improve the value function, the parameters $\theta_v$ are updated to minimize the $\ell^2$ error $\mathbb{E}\big(R_{t} - \approxval{t}\big)^2$ between the value function, $\approxval{t}$, and the observed return, $R_{t}$.  The observed return is obtained by using the current policy to select actions to apply to the reaction simulation environment.

The training therefore proceeds iteratively, with the current value function being used to update the policy and the current policy being used to update the value function.  The parameter updates occur periodically throughout the course of an episode, or single polymerization reaction.  The current policy is first used to generate a {length-$L$} sequence of state transitions ${\{s_{t},\,a_{t},\,r_{t},\,s_{t+1},\,a_{t+1},\,r_{t+1},\,\cdots,\,s_{t+L}\}}$.  This {length-$L$} sequence is referred to as a \emph{rollout}.  At the end of each rollout, the information generated during the rollout is used to update $\theta_p$ and $\theta_v$.  To take advantage of multi-core computing architectures, the training process is distributed to multiple asynchronous parallel learners.  A3C keeps a global version of $\theta_p$ and $\theta_v$.  Each learner has access to a separate copy of the reaction simulation environment and a local version of $\theta_p$ and $\theta_v$.  After a learner performs a rollout, it generates updates to $\theta_p$ and $\theta_v$.  These updates are then applied to the global versions of $\theta_p$ and $\theta_v$, and the learner replaces its local version with the global version.  In this manner, each learner periodically incorporates updates generated by all learners.

\subsection{\label{ssec:imple}Additional implementation details}
The neural networks used for the policy and value functions share a common stack of hidden layers, but use separate final output layers.  We compare results from two different network architectures for the hidden layers.  The first architecture, FCNN, is a simple fully-connected neural network with two hidden layers containing 200 and 100 hidden units, respectively.  The second architecture, 1D-CNN, is convolutional.  In 1D-CNN, the input feature vector is fed into a first 1D convolutional layer having 8 filters of length 32 with stride 2, followed by a second 1D convolutional layer having 8 filters of length 32 with stride 1.  The output of the second 1D convolutional layer is then fed into a fully-connected layer with 100 units.  All hidden layers use rectifier activation.  The final layer of the value network produces a single scalar output that is linear in the 100 units of the last hidden layer.  The final layer of the policy network is a softmax layer of the same 100 hidden units, with a length-6 output representing a probability distribution over the 6 actions.  For a crude estimate of model complexity, FCNN and 1D-CNN contain 42607 and 9527 trainable parameters, respectively.

We implemented the A3C algorithm with 12 parallel CPU learners.\cite{mnih2016asynchronous}  The discount factor in the return is $\gamma = 0.99$, and the maximum rollout length is $20$.  The length of a rollout may be shorter than 20 when the last state in the sequence is a terminal state.  After a learner collects a {length-$L$} rollout, ${\{s_{t},\,a_{t},\,r_{t},\,s_{t+1},\,a_{t+1},\,r_{t+1},\,\cdots,\,s_{t+L}\}}$, it generates updates for $\theta_p$ and $\theta_v$ by performing stochastic gradient descent steps for each ${t'\in\{t,\,\cdots,\,t+L-1\}}$.  Define the bootstrapped multi-step return ${R_{t'}'=\indic\gamma^{t+L-t'}\approxvalloc{t+L} +  \sum_{i=t'}^{t+L}\gamma^{i-t'}r_i}$ where ${\indic = 0}$ if $s_{t+L}$ is the terminal state and $1$ otherwise.  The prime on $\theta_v'$ in $\approxvalloc{t+L}$ indicates that the value function is evaluated using the local copy of the network parameters.  The update direction of $\theta_p$ is set according to
\[d\theta_p = -\nabla_{\theta_p'}\log\locpola{t'}\big(R_{t'}' - \approxvalloc{t'}\big) + \beta\nabla_{\theta_p'}\entropy.\]
$\entropy$ is the entropy of $\locpol{t'}$ and acts as a regularization term that helps prevent $\locpol{t'}$ from converging to sub-optimal solutions.  $\beta$ is the regularization hyperparameter, for which we use $\beta = 0.01$.  $\theta_v$ is updated according to the direction of
\[d\theta_v = \nabla_{\theta_v'}\big(R_{t'}' - \approxvalloc{t'}\big)^2.\]
Updates of the network parameters are done using the ADAM optimizer\cite{kingma2014adam} with a learning rate of $1\times10^{-4}$.

Additionally, after each action is drawn from the probability distribution generated by the policy, the agent repeats the action for 4 times before selecting the next action.  This repetition shortens the length of a full episode by a factor of 4 from the RL agent's perspective and so prevents the value function from exponential vanishing.\cite{mnih2013playing}

\section{Results and discussion}

\subsection{Targeting Gaussian MWDs with different variance}
Our first goal is to train the RL controller against some MWDs with simple analytic forms, for which Gaussian distributions with different variances seem a natural choice.  Seemingly simple, Gaussian MWDs exemplify the set of symmetric MWDs the synthesis of which requires advanced ATRP techniques such as activators regenerated by electron transfer (ARGET).\cite{listak2008effect}  Living polymerization produces a Poisson distribution with a variance that depends only on the average chain length, which is set by the monomer-to-initiator ratio.  The variance from the ideal living polymerization provides a lower limit to the variance of the MWD.  Here, we choose Gaussian distributions with variances ranging from near this lower limit to about twice that limit.  Increasing the variance of the MWD can have substantial effects on the properties of the resulting material.\cite{lynd2008polydispersity}

\begin{figure}[ht]
\centering
\includegraphics[width=0.95\columnwidth]{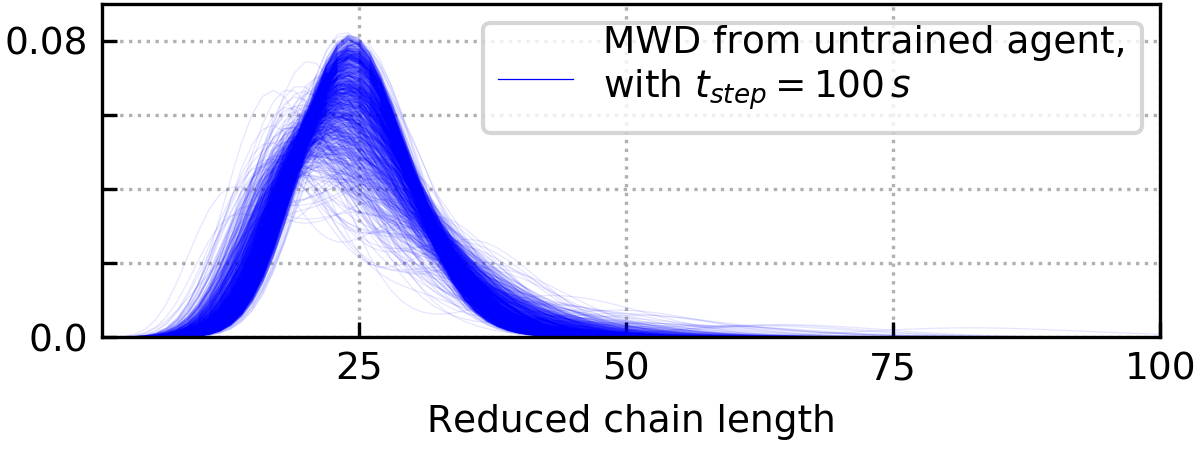}
\caption{Superposition of 1000 ending MWDs from untrained agents when the time interval between actions is 100 seconds. Vertical axis is fraction of polymer chains.}
\label{fig:gv_untrained}
\end{figure}

\begin{figure}[ht]
\centering
\includegraphics[width=0.95\columnwidth]{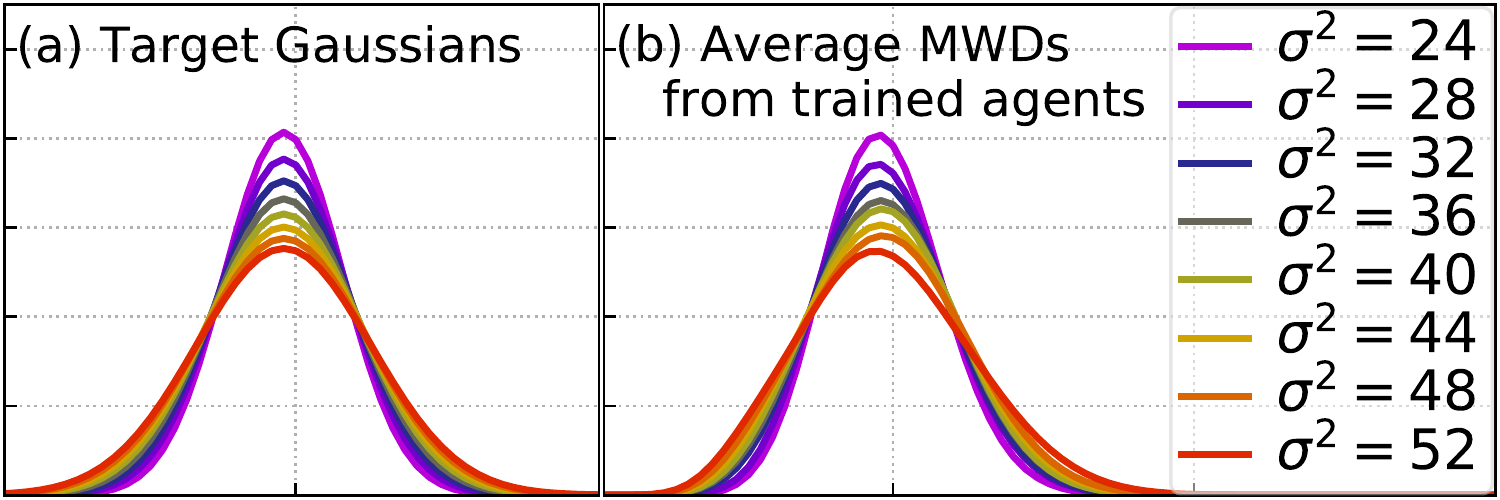}
\caption{Comparison of the human-specified target Gaussian MWDs with the average ending MWDs given by trained 1D-CNN agents, with averaging being over 100 episodes.  The horizontal and vertical spacings between dotted line grids are 25 and 0.02, respectively.}
\label{fig:gv_targets_effects}
\end{figure}

For this task, we set the time interval between two actions to 100 seconds.  This setting was chosen for two main reasons.  First, due to the choice of the addition unit amounts and budget limits of addable reagents, it typically takes $300{\sim}400$ simulator steps to finish one episode, and so this choice of time interval corresponds to ${\sim}10$ hours of real reaction time before the terminal step.  More importantly, it allows an untrained RL controller to produce a widely variable ending MWD, as illustrated by the 1000 MWDs of \fig~\ref{fig:gv_untrained}.  A widely variable ending MWD is necessary for RL agents to discover strategies for target MWDs through self-exploration.\cite{jaksch2010near,kearns2002near}

As specific training targets, we select Gaussian MWDs with variances ($\sigma^2$'s) ranging from 24 to 52, which covers the theoretical lower limit of the variance to a variance of more than twice this limit.  \fig~\ref{fig:gv_targets_effects}(a) shows the span of these target MWDs.  A summary of the trained 1D-CNN agents' performance on this task is shown in \fig~\ref{fig:gv_targets_effects}(b).  Each ending MWD is an average over 100 episodes, generated using the trained 1D-CNN controller.  Note that this MWD averaging is equivalent to blending polymer products generated in different reactions,\cite{leonardi2017automatic} a common practice in both laboratory and industrial polymerization.\cite{jovanovic2004butyl,lenzi2005producing,deslauriers2005comparative,zhang2002modeling}  The trained 1D-CNN agent used in these test runs is that which gave the best performance in the training process, i.e., the neural network weights are those that generated the highest reward during the training process.  During training, termination reactions are not included in the simulation, but during testing, these reactions are included.  For all 8 target Gaussian MWDs, the average ending MWDs are remarkably close to the corresponding targets.  The maximum absolute deviation from the target MWD is an order of magnitude less than the peak value of the distribution function.  These results show that control policies learned on simulation environments that exclude termination transfer well to environments that include termination.  This is perhaps not surprising because ATRP of styrene is close to an ideal living polymerization, with less than 1\% of monomers residing in chains that underwent a termination reaction.  Tests on changing other aspects of the polymerization simulation are given in the following sections.

\newcommand{\onewidth}{0.48\columnwidth}
\begin{figure}[b!]
  \centering
    \includegraphics[width=\onewidth]{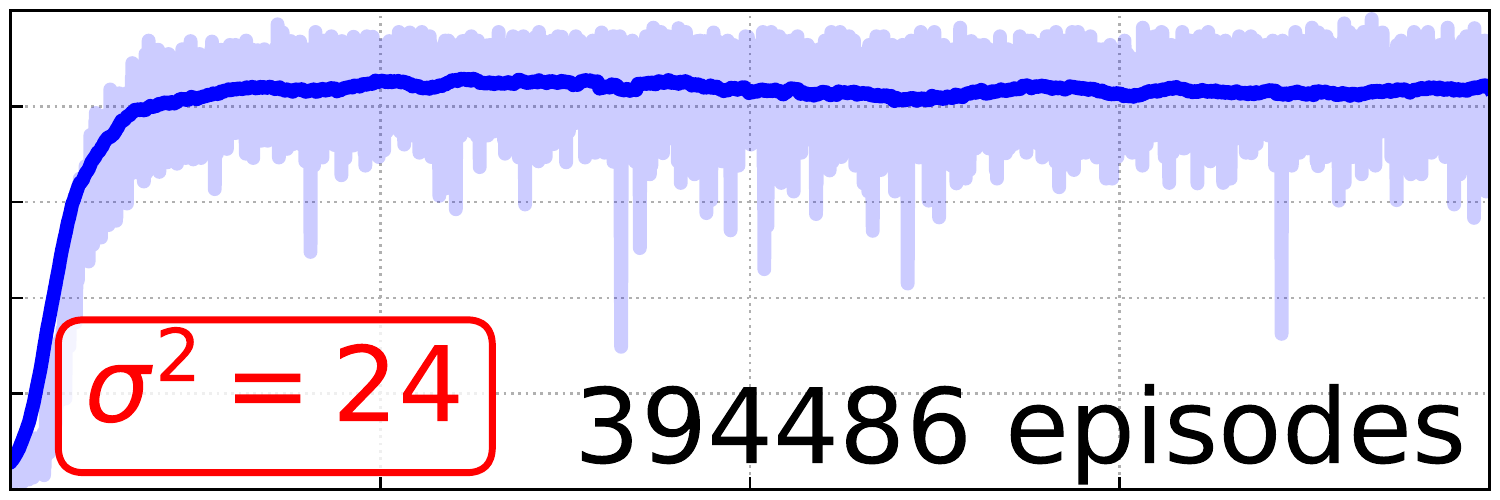}
    \includegraphics[width=\onewidth]{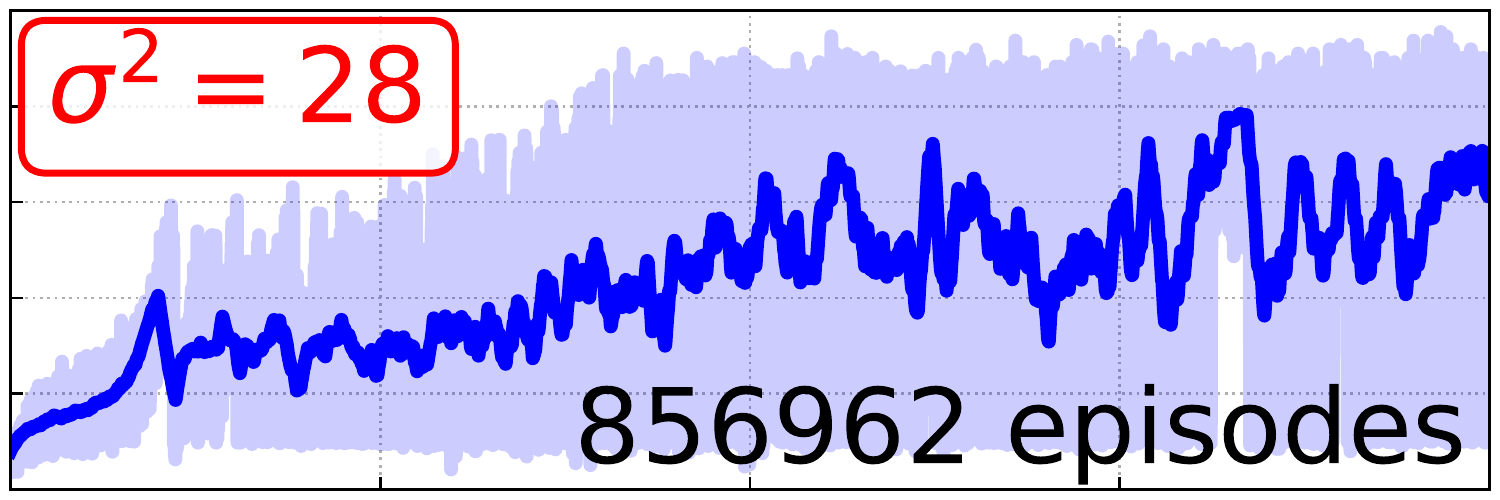} \\
    \includegraphics[width=\onewidth]{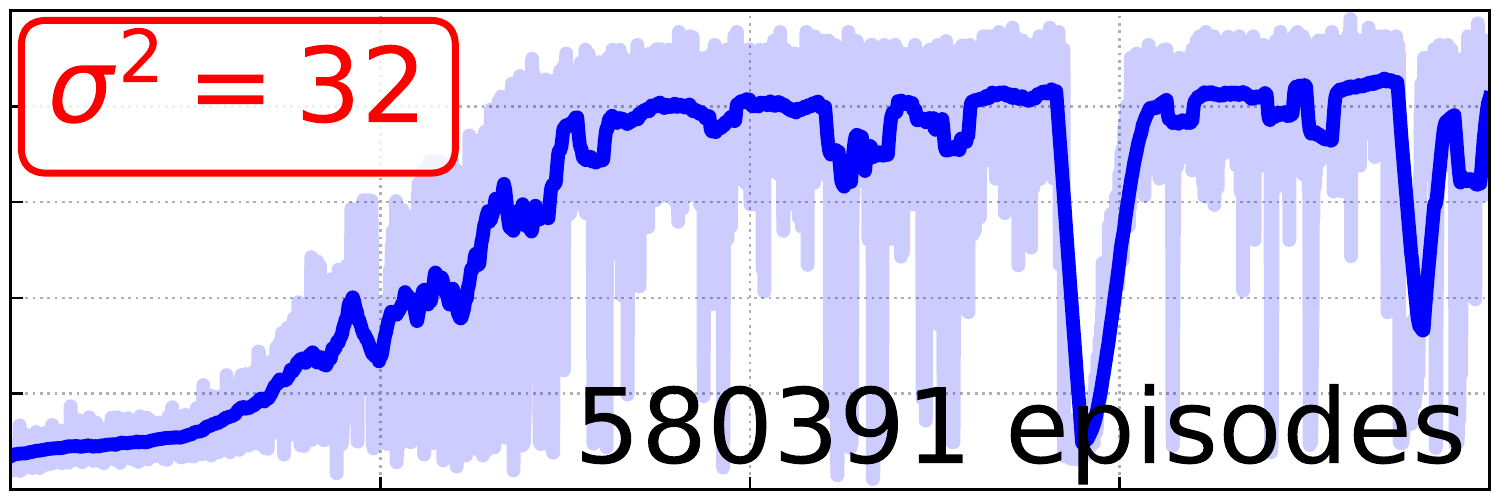}
    \includegraphics[width=\onewidth]{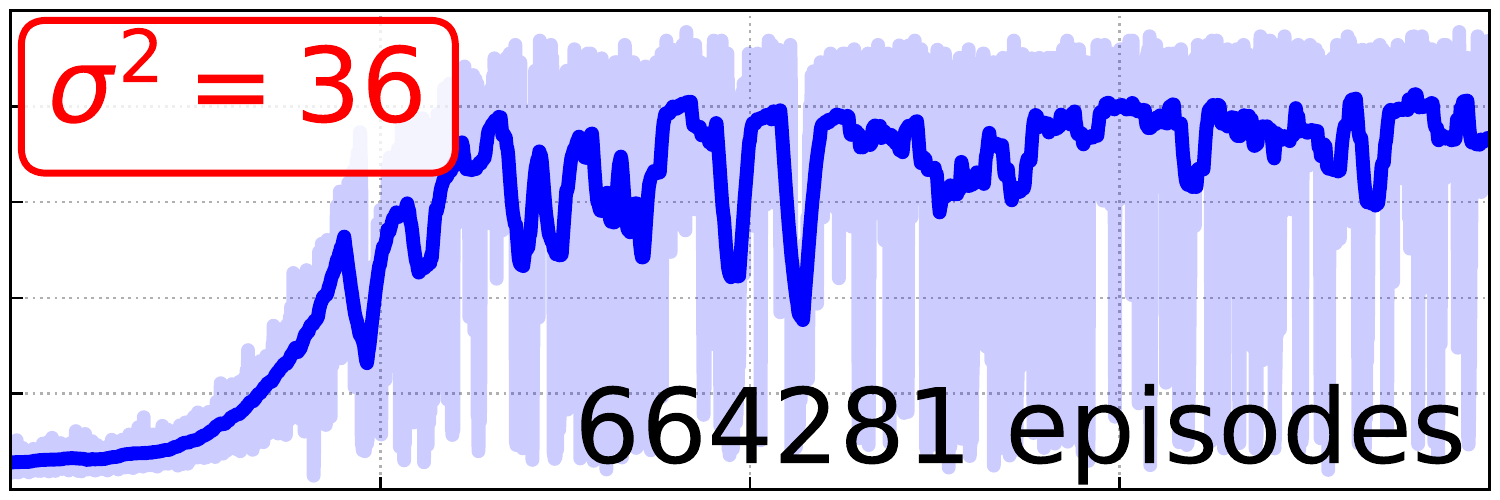} \\
    \includegraphics[width=\onewidth]{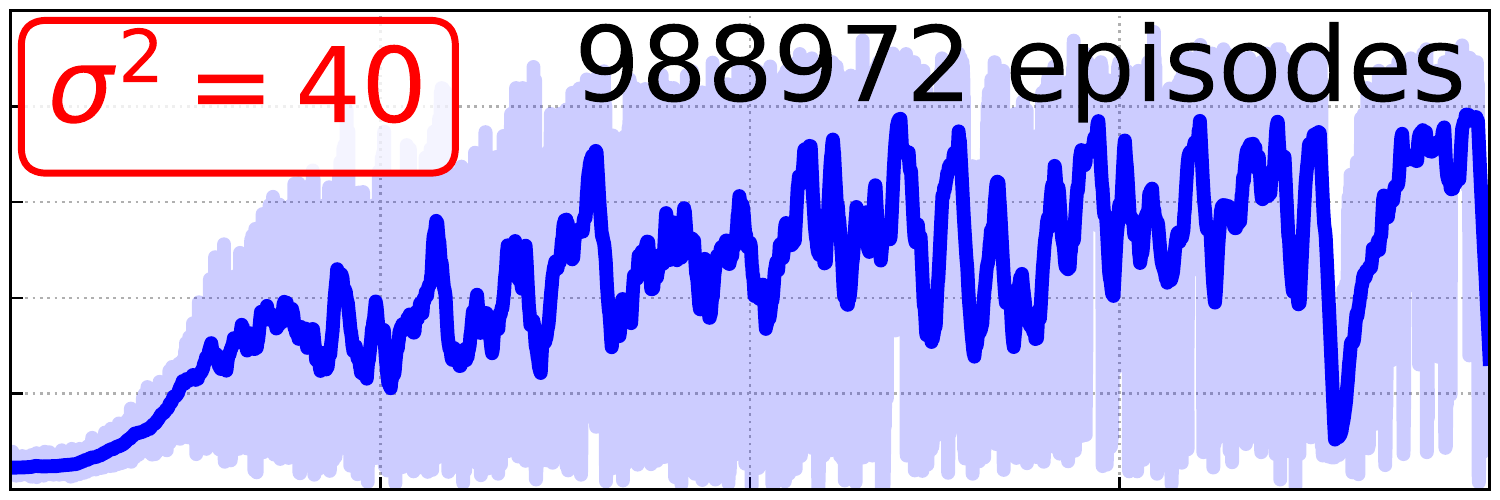}
    \includegraphics[width=\onewidth]{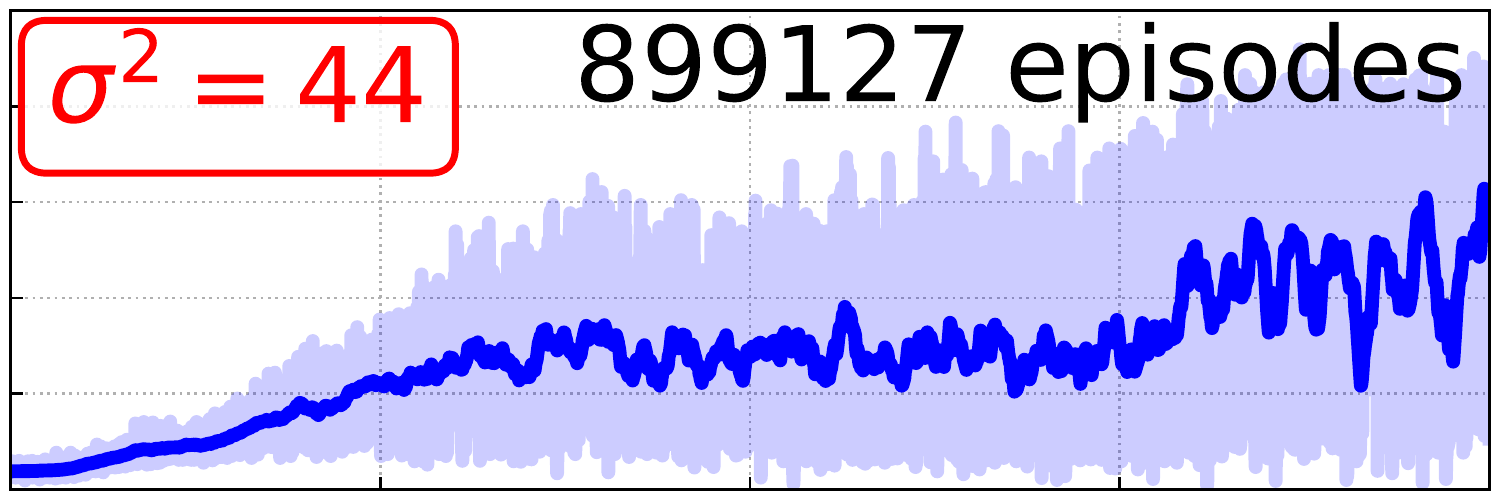} \\
    \includegraphics[width=\onewidth]{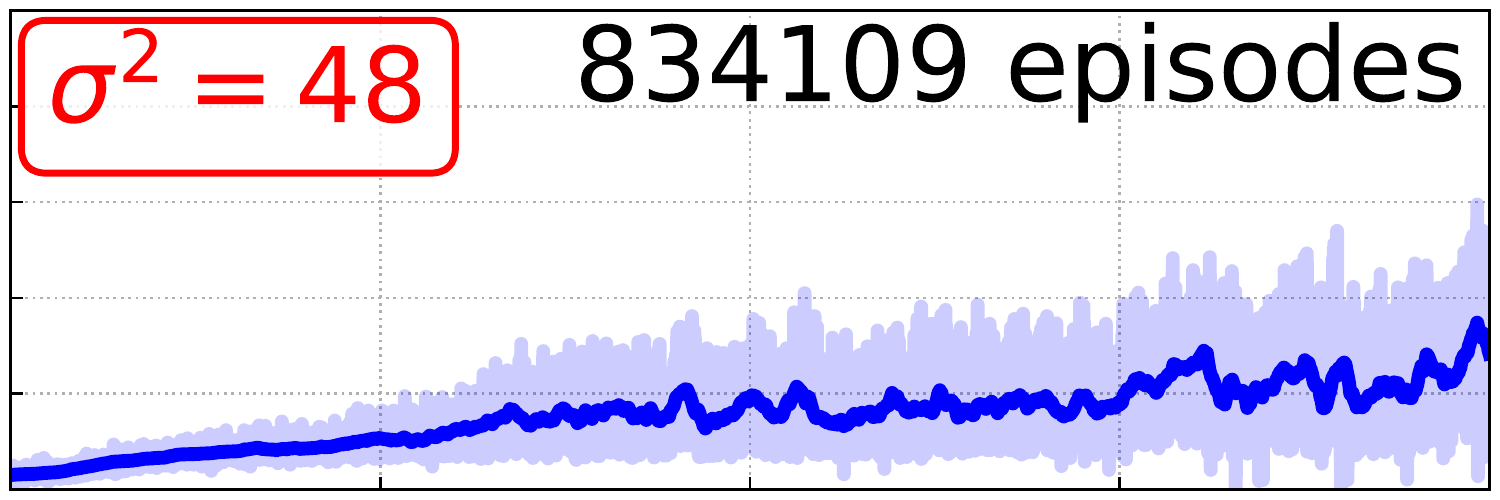}
    \includegraphics[width=\onewidth]{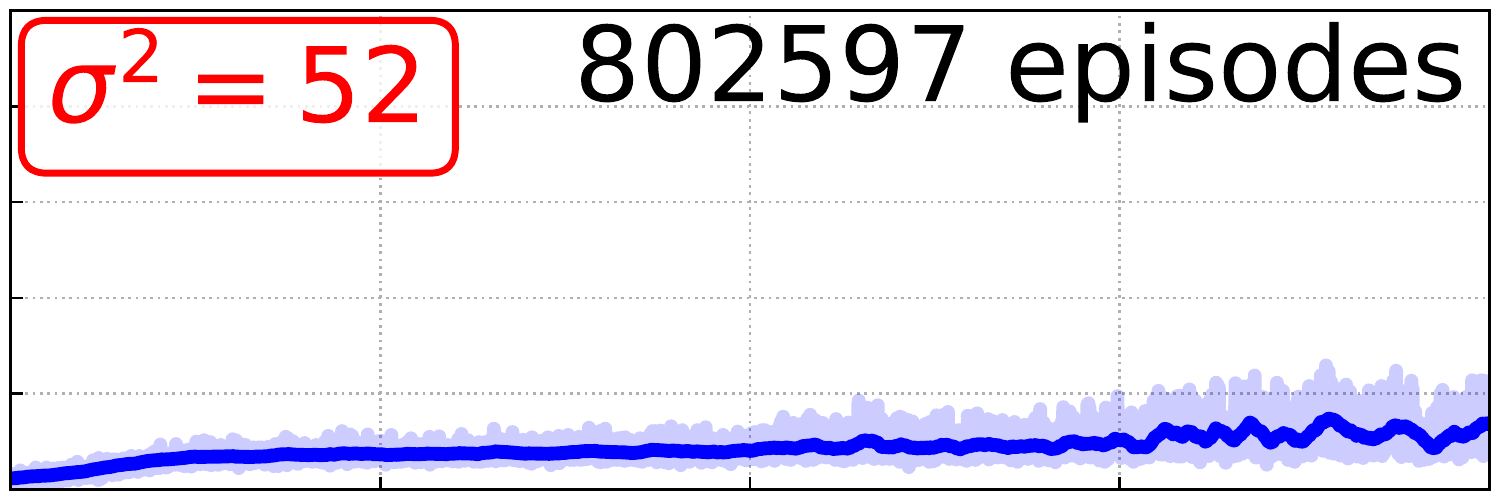}
  \caption{Learning curves for training FCNN agents on the target Gaussian MWDs of \fig~\ref{fig:gv_targets_effects}.  Horizontal axis is number of episodes, or simulated reactions, with total number of episodes shown in legend.  The vertical axis is the instantaneous (light blue) or averaged (dark blue) reward, as defined in the main text, on a 0 to 1 scale.}
  \label{fig:learning_fcnn}
\end{figure}

\begin{figure}[b!]
  \centering
    \includegraphics[width=\onewidth]{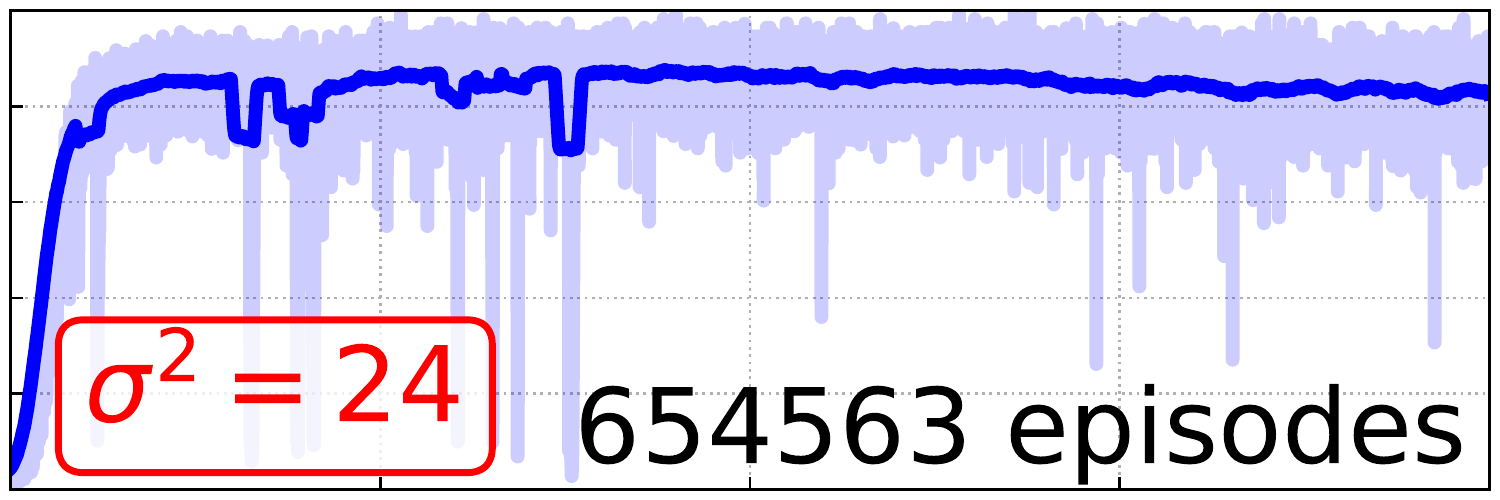}
    \includegraphics[width=\onewidth]{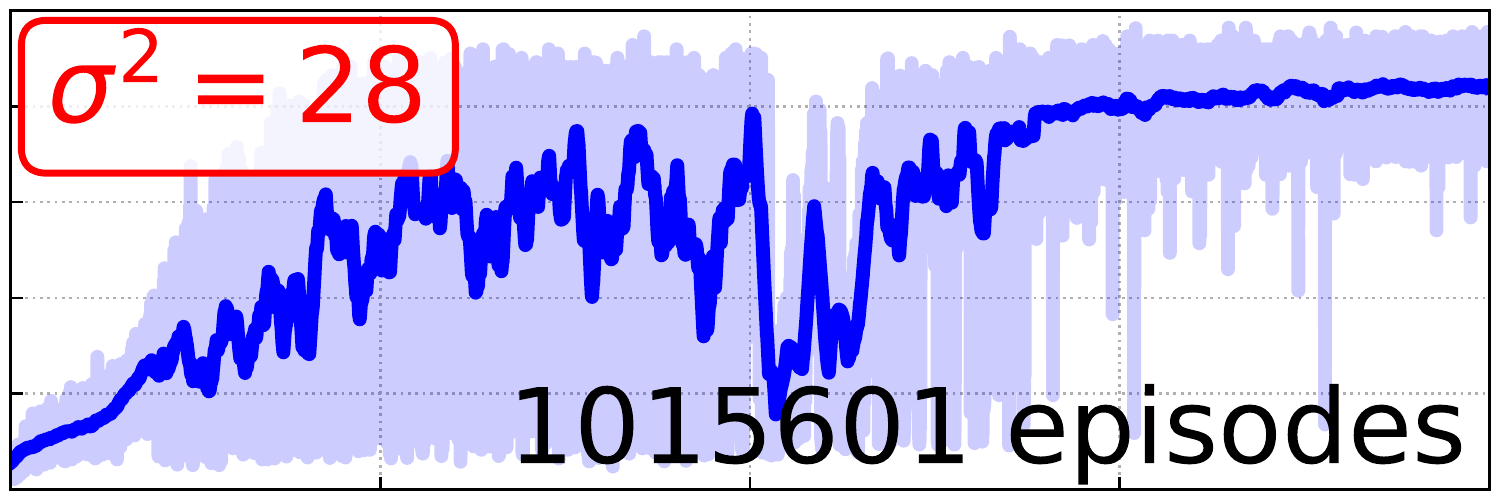} \\
    \includegraphics[width=\onewidth]{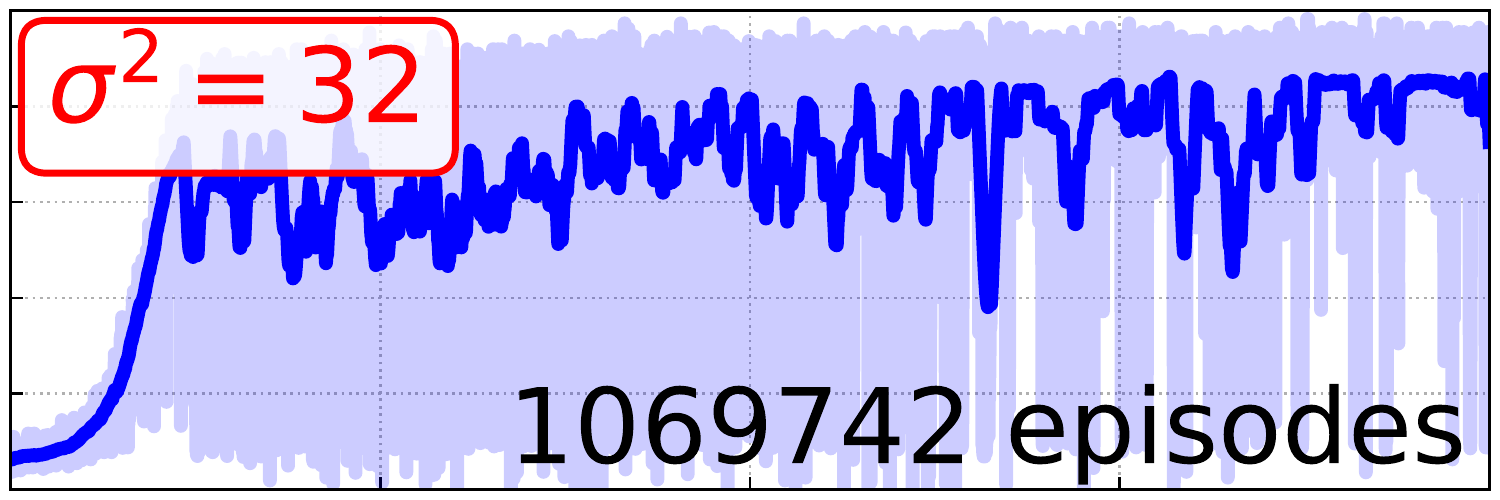}
    \includegraphics[width=\onewidth]{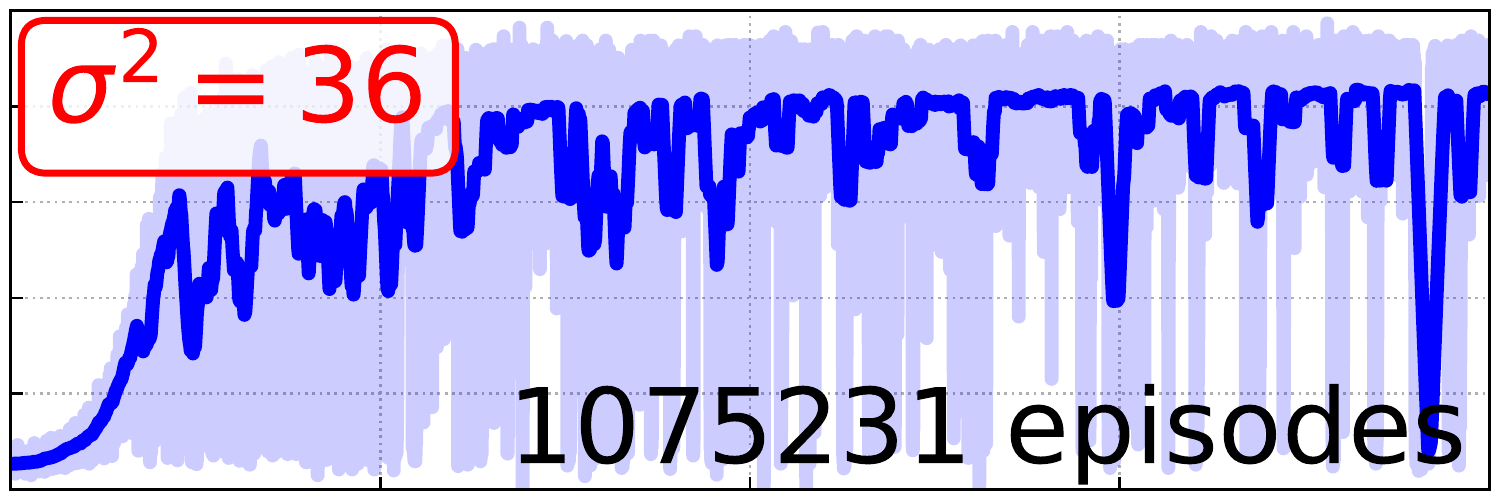} \\
    \includegraphics[width=\onewidth]{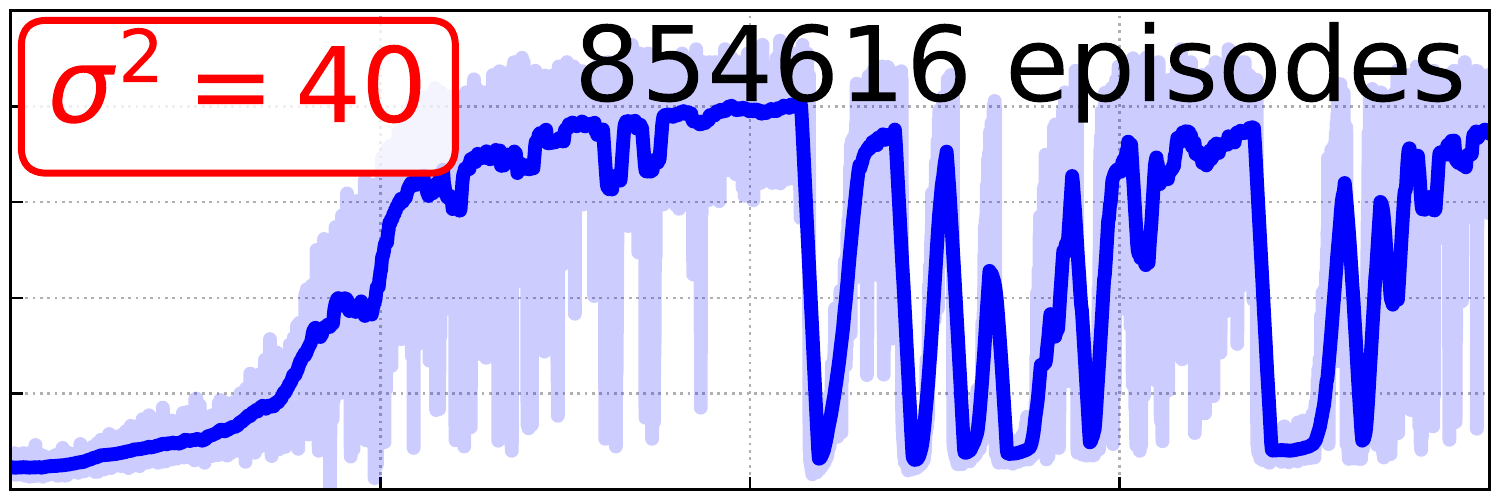}
    \includegraphics[width=\onewidth]{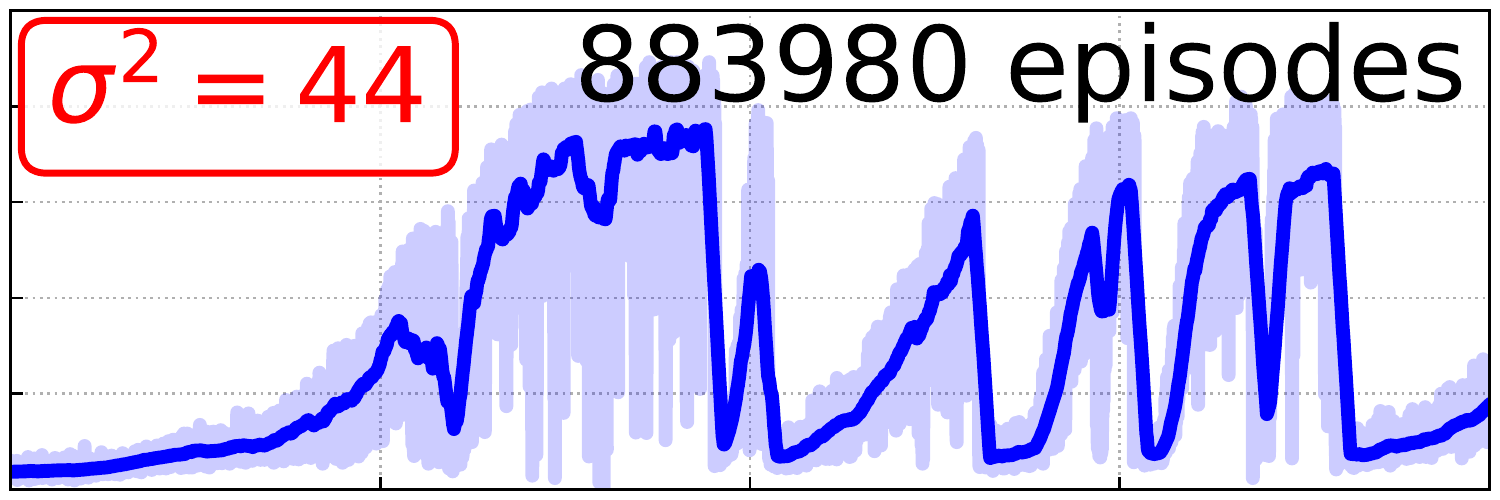} \\
    \includegraphics[width=\onewidth]{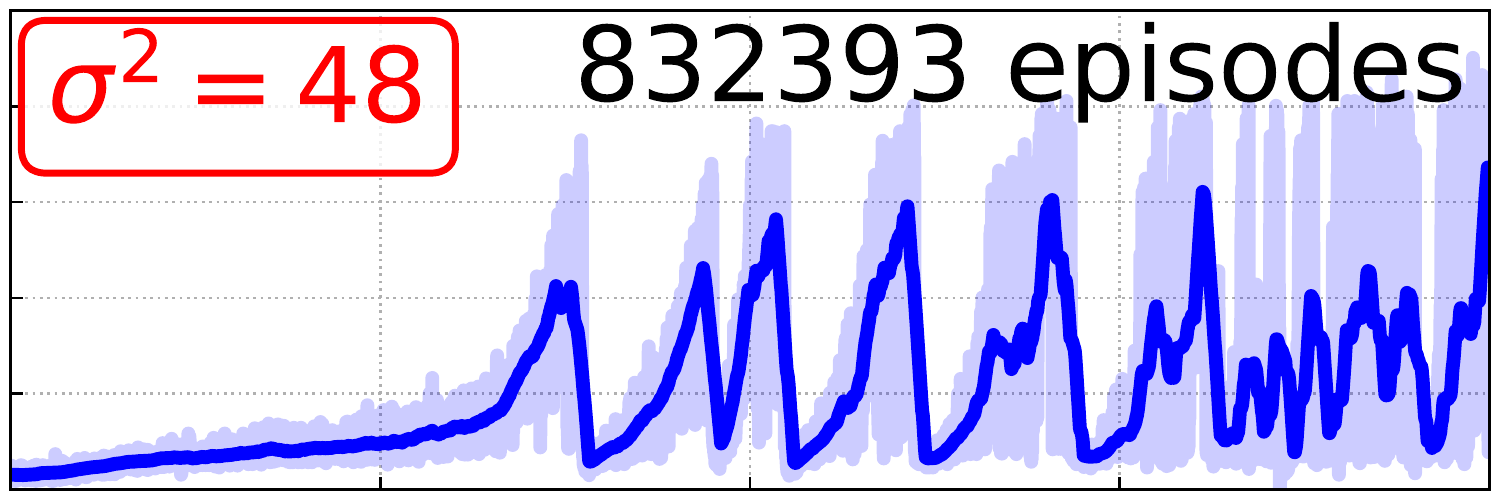}
    \includegraphics[width=\onewidth]{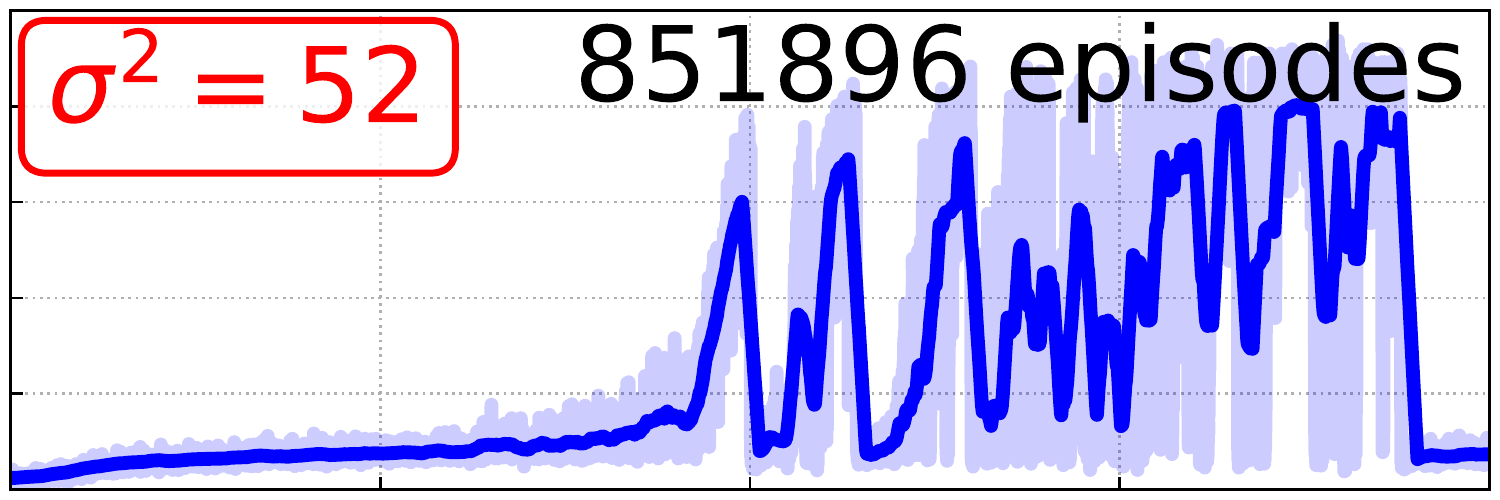}
  \caption{Learning curves for training 1D-CNN agents on the target Gaussian MWDs of \fig~\ref{fig:gv_targets_effects}. Convention is as in \fig~\ref{fig:learning_fcnn}.}
  \label{fig:learning_1dcnn}
\end{figure}

\subsubsection{Training process and learning curves}

\fig~\ref{fig:learning_fcnn} and \ref{fig:learning_1dcnn} compare the learning curves of FCNN and 1D-CNN agents.  The horizontal axis shows the number of ATRP experiments (episodes) run by the agent during the training process.  The vertical axis shows the reward received by the agent, which runs from 0.0 to the maximum possible reward of 1.0.  The dark blue lines average over a window of length 10000 and so reflect the agents' average performance during training.  The light blue regions average over a window of length 100 and so may be interpreted as the agents' instantaneous performance.  The transition from low to high average reward partially reflects the two-level reward structure, in which the reward is 0.1 for loose agreement with the target MWD and 1.0 for tight agreement.

\newcommand{\effwidth}{0.45\columnwidth}
\begin{figure*}[t]
  \centering
  \includegraphics[width=\effwidth]{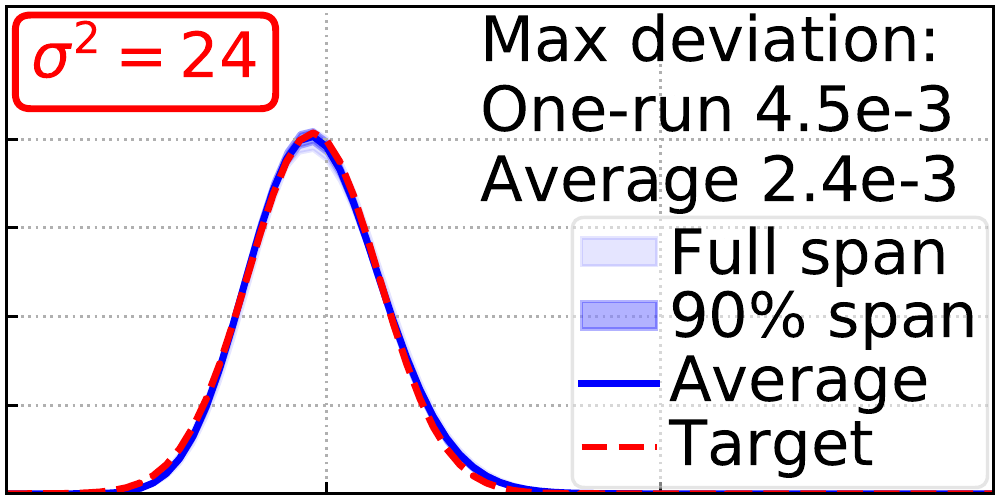}
  \includegraphics[width=\effwidth]{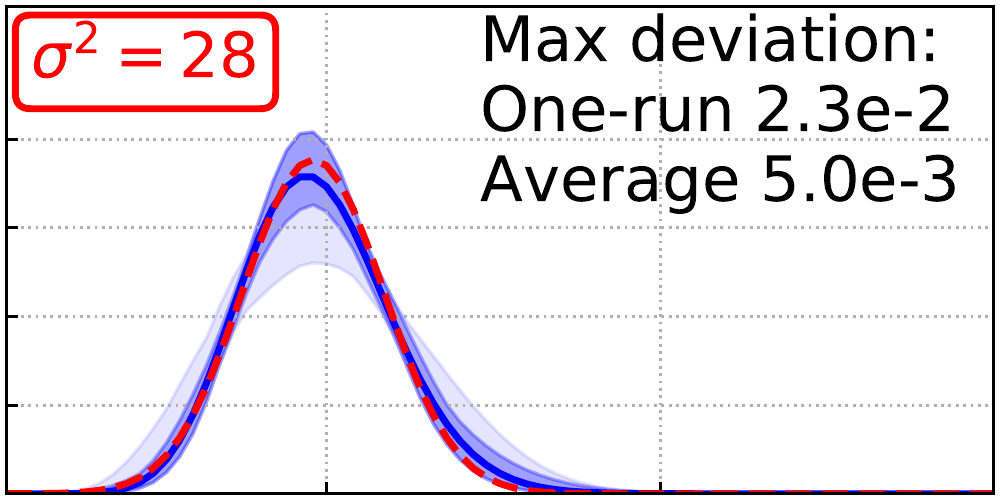}
  \includegraphics[width=\effwidth]{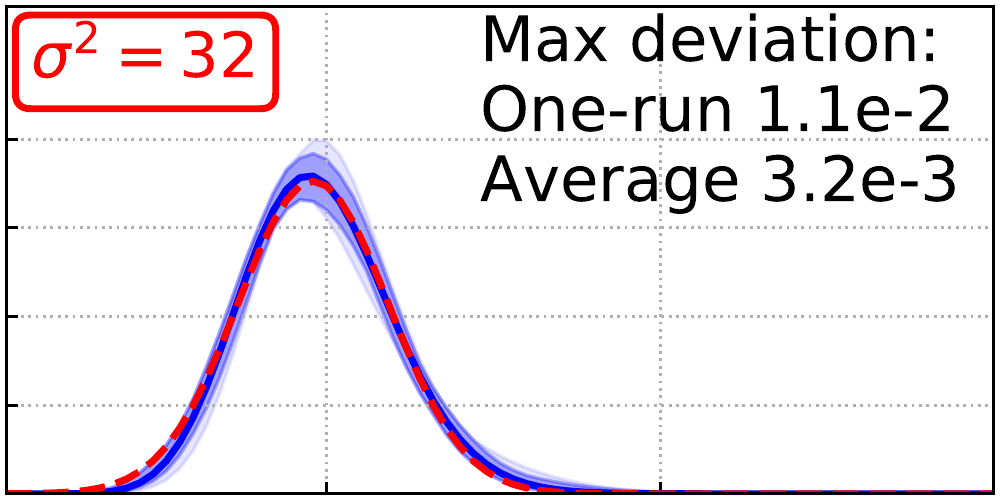}
  \includegraphics[width=\effwidth]{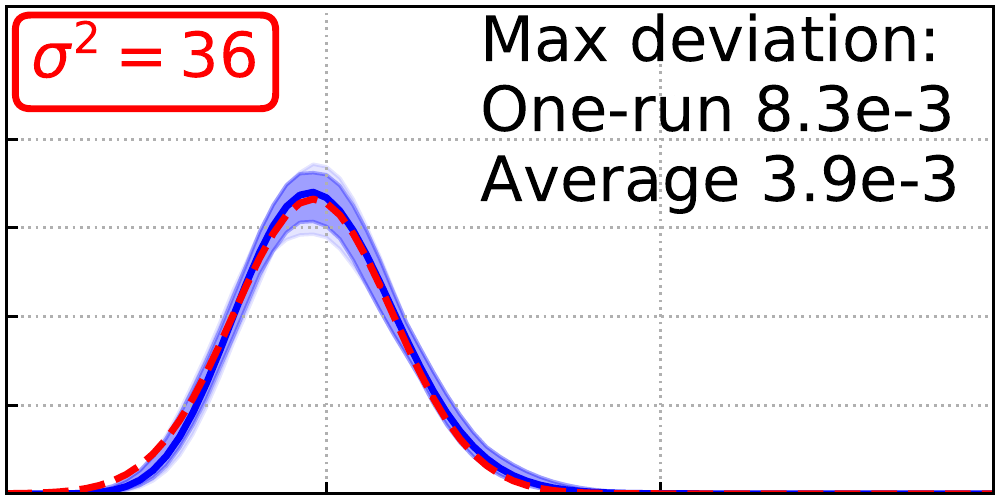} \\
  \includegraphics[width=\effwidth]{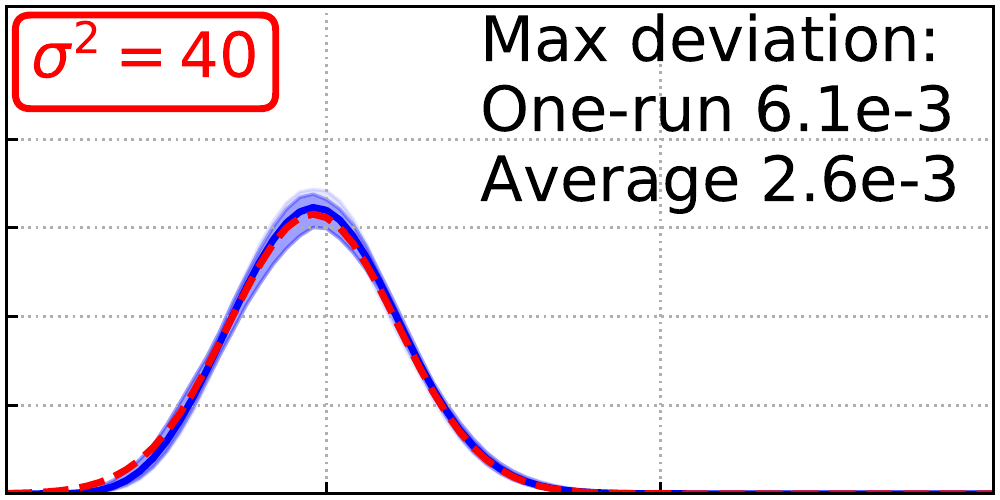}
  \includegraphics[width=\effwidth]{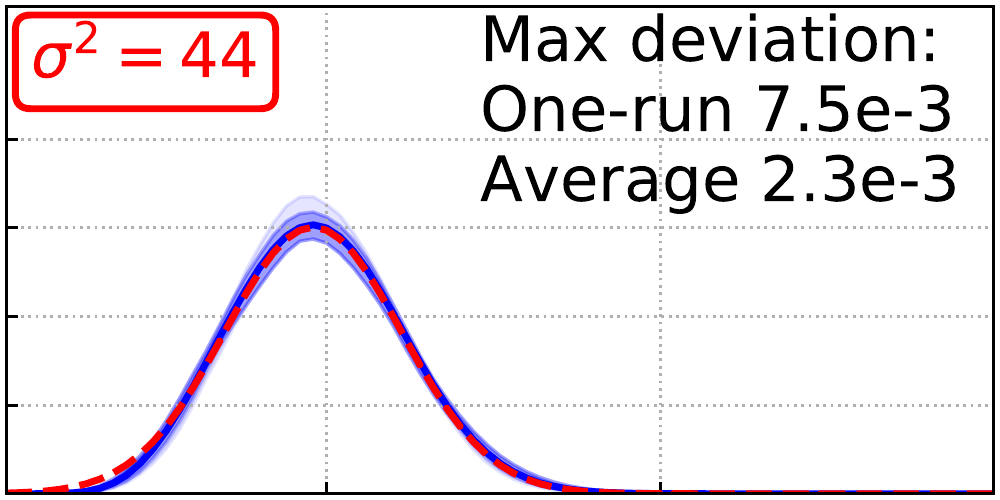}
  \includegraphics[width=\effwidth]{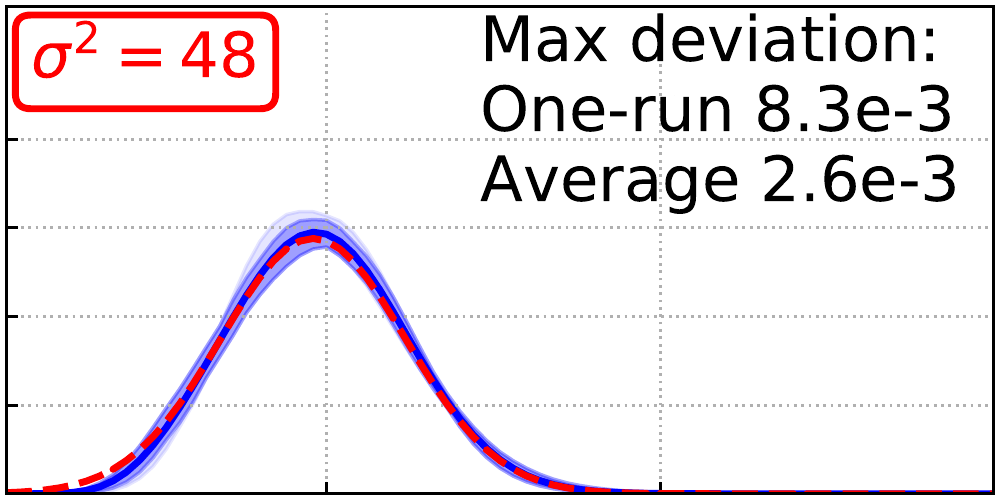}
  \includegraphics[width=\effwidth]{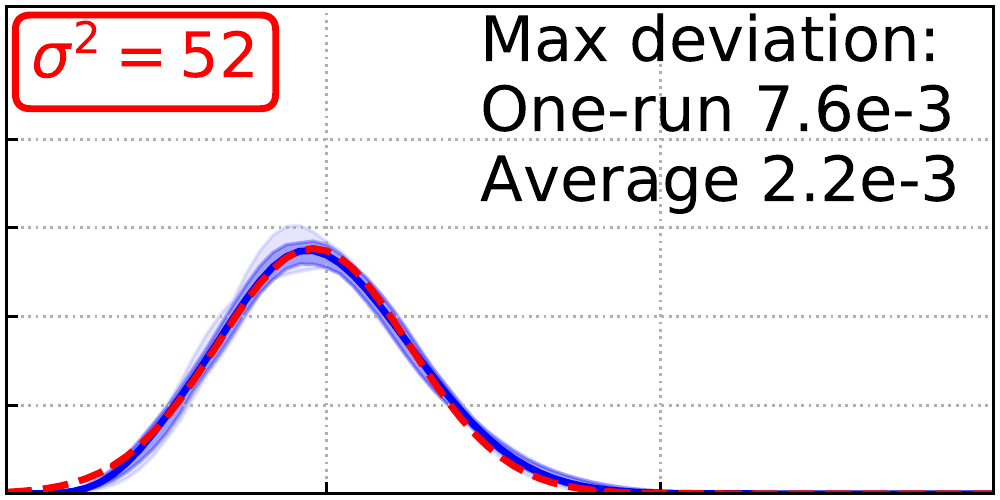}
  \caption{Performance of 1D-CNN agents trained on the target Gaussian MWDs of \fig~\ref{fig:gv_targets_effects} on simulation environments that include both termination reactions and noise.  In each subplot, the horizontal axis represents the reduced chain length and runs from 1 to 75, and the vertical axis represents fraction of polymer chains and runs from 0.0 to 0.11.}
  \label{fig:gv_transfer_effect}
\end{figure*}

\begin{figure*}[t]
  \centering
  \includegraphics[width=\effwidth]{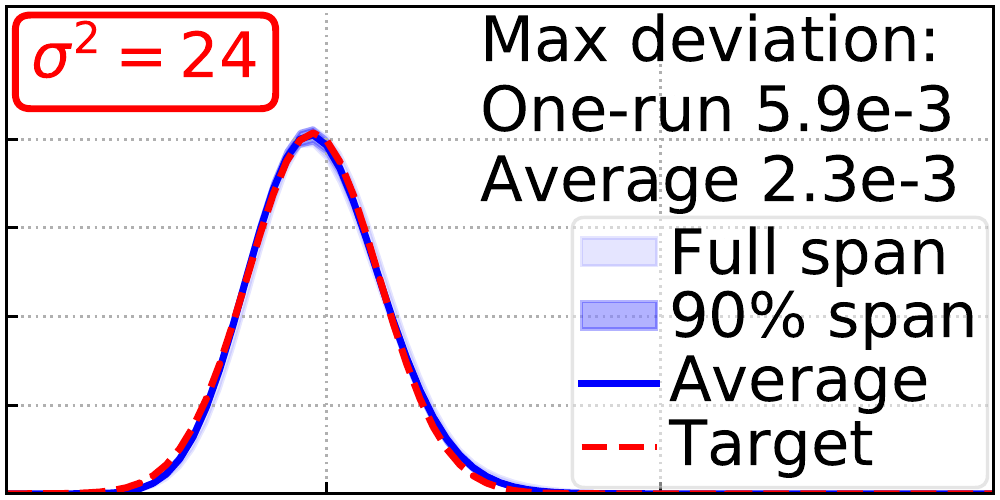}
  \includegraphics[width=\effwidth]{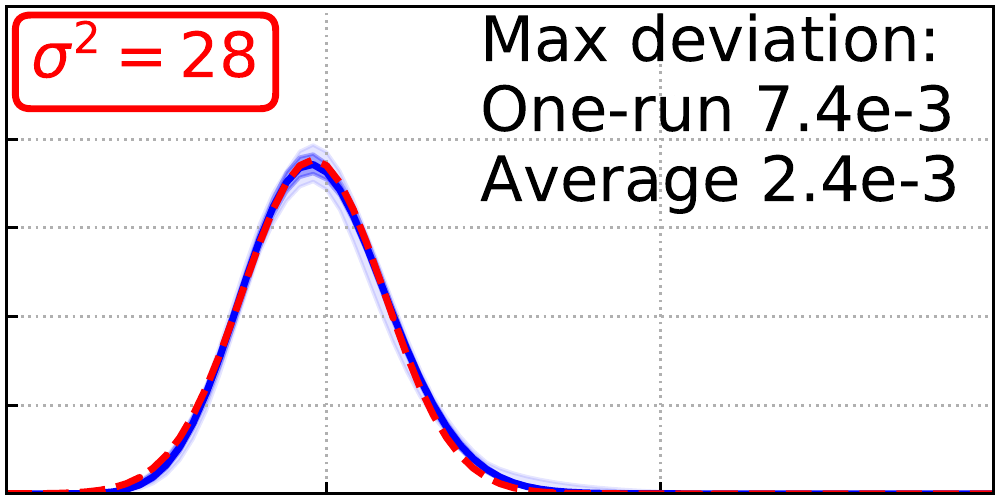}
  \includegraphics[width=\effwidth]{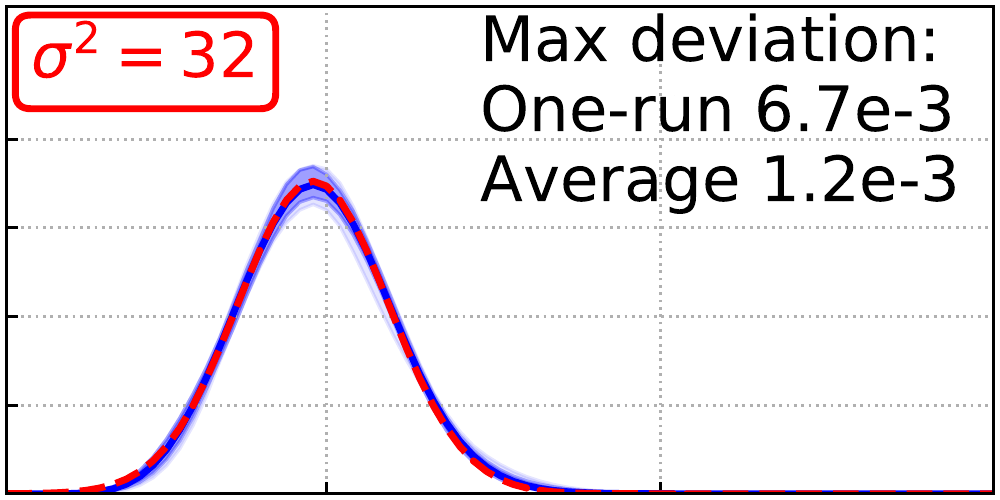}
  \includegraphics[width=\effwidth]{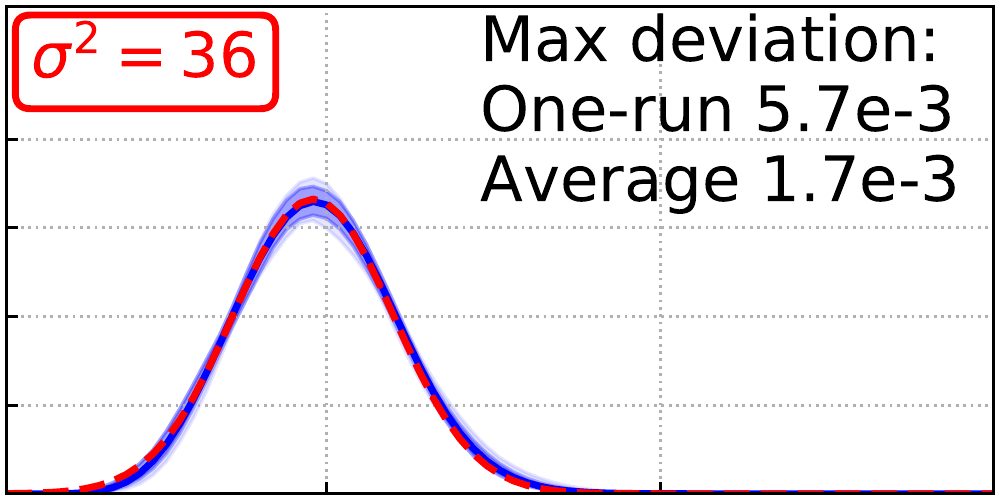} \\
  \includegraphics[width=\effwidth]{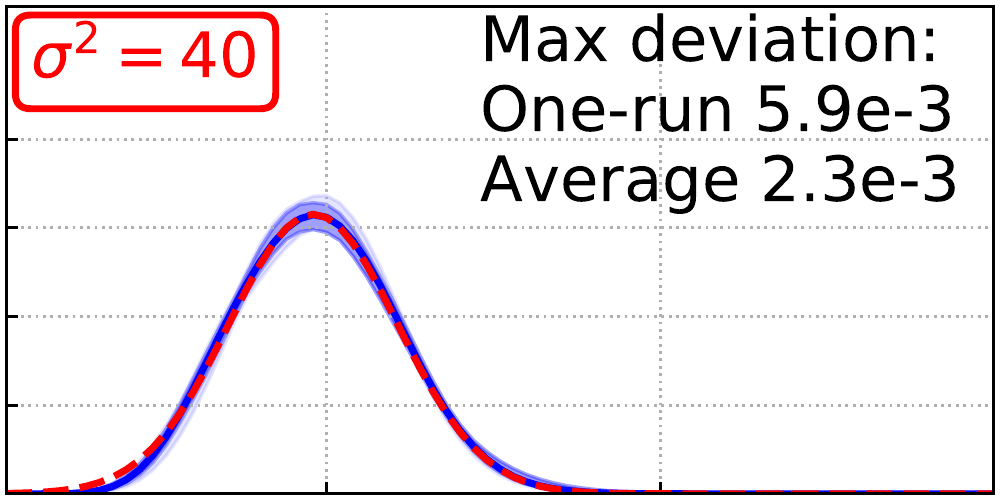}
  \includegraphics[width=\effwidth]{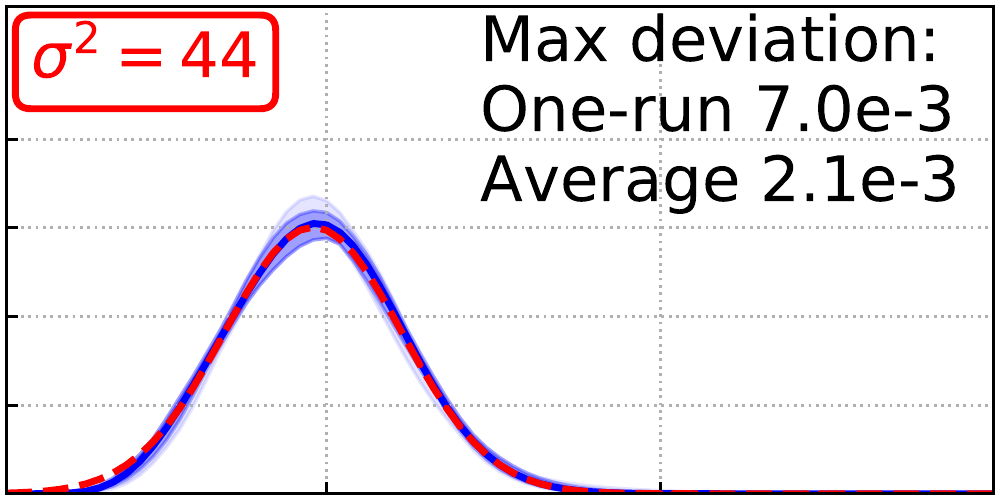}
  \includegraphics[width=\effwidth]{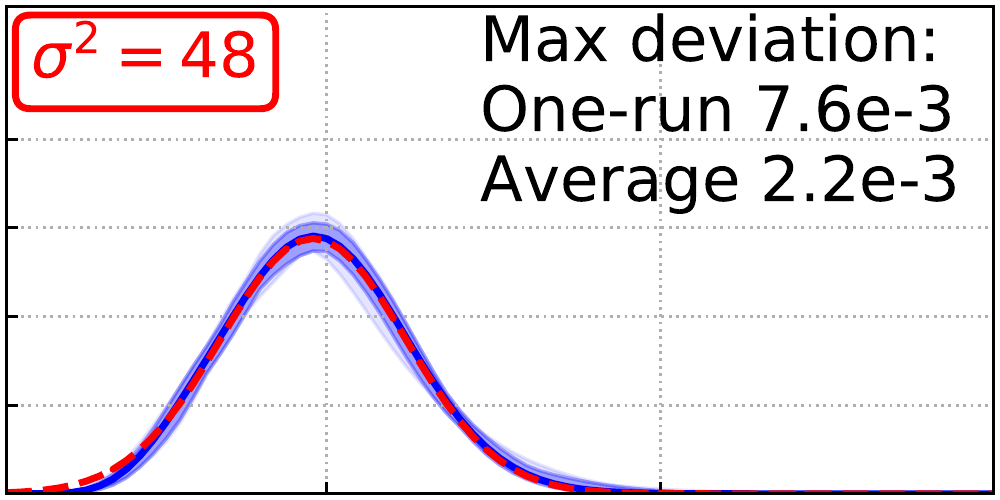}
  \includegraphics[width=\effwidth]{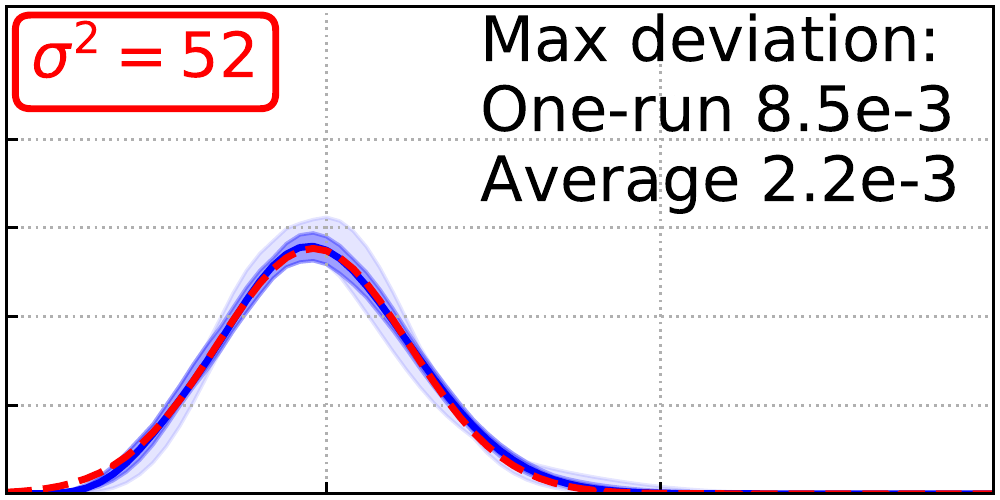}
  \caption{Performance of 1D-CNN agents trained on noisy, with-termination environments targeting Gaussian MWDs of \fig~\ref{fig:gv_targets_effects}. Convention is as in \fig~\ref{fig:gv_transfer_effect}.}
  \label{fig:gv_srcg_effect}
\end{figure*}

Two general trends emerge from these learning curves.  The first is that broader target MWDs require strategies that are harder for the RL agents to learn.  When the target MWD is a Gaussian with variance 24, both FCNN and 1D-CNN can learn a strategy in less than $10^5$ training episodes.  As the variance of the target distribution increases, the number of required training episodes increases substantially.  The second general trend is that the 1D-CNN outperforms the FCNN.  For the narrower target distributions, both architectures obtain similar peak performance, but the 1D-CNN trains faster and the performance is more steady.  For broader target distributions, only the 1D-CNN could achieve the 1.0 tight-threshold reward consistently.

\subsubsection{Transferability tests on noisy environments}
To test the robustness of the learned control policies, the trained 1D-CNN agents were evaluated on simulation environments that include both termination reactions and simulated noise.\cite{duan2016benchmarking,hester2013open,bakker2002reinforcement}  We introduce noise on the states as well as actions.  On states, we apply Gaussian noise with standard deviation $1\times10^{-3}$ on every observable quantity.  (The magnitude of the observable quantities range from 0.01 to 0.1.) In the simulation, we introduce three types of noise.  First, the time interval between consecutive actions is subject to a Gaussian noise, whose standard deviation is 1\% of the mean time interval.  Gaussian noise is also applied to the amount of chemical reagent added for an action, again with a standard deviation that is 1\% of the addition amount.  Lastly, every kinetics rate constant used in non-terminal steps is subject to Gaussian noise, with the standard deviation being 10\% of the mean value.  Note that we crop the Gaussian noise in the simulation at $\pm3\sigma$ to avoid unrealistic physics, such as negative time intervals, addition of negative amounts, or negative kinetic rate constants.  Once all budgets have been met, the simulation enters its terminal step and the RL agent no longer has control over the process.  During this terminal step, we do not apply noise.

Performance of the 1D-CNN agents, trained against the target Gaussian MWDs of \fig~\ref{fig:gv_targets_effects}, on noisy environments is shown in \fig~\ref{fig:gv_transfer_effect}.  The trained agent is used to generate 100 episodes and the statistics of final MWDs are reported in a variety of ways.  The average MWD from the episodes is shown as a solid dark blue line.  The light blue band shows the full range of the 100 MWDs and the blue band shows, at each degree of polymerization, the range within which 90 of the MWDs reside.  The control policies learned by 1D-CNN agents seem to be robust.  The deviation of the average MWD is an order of magnitude less than the peak value of the MWD.  Deviations of the MWD from a single episode can vary more substantially from the target MWD, but the resulting MWDs are still reasonably close to the target MWD.  On average, the maximum absolute deviation between a one-run MWD and the target is still less than 5\% of the peak MWD value.

\subsubsection{Training directly on noisy environments}
Training the RL agents on noisy environments can significantly reduce the deviations of the single-run MWDs from the target MWD, as shown in \fig~\ref{fig:gv_srcg_effect}.  Noticeably, on the $\sigma^2 = 28$ environment, the ``one-run'' maximum absolute deviation is reduced from $2.3\times10^{-2}$ to $7.4\times10^{-3}$, a reduction of over a factor of 3.  These results are consistent with an expected advantage of state-dependent control policies, where the agents can respond to the real-time status of the reactor and autonomously choose the proper action to perform.  Even though the states may be noisy, the RL agents are still able to detect patterns and use them to form a probability distribution over actions that maximizes the chance of reaching the target MWDs.

\begin{figure}[ht]
  \centering
  \includegraphics[width=0.95\columnwidth]{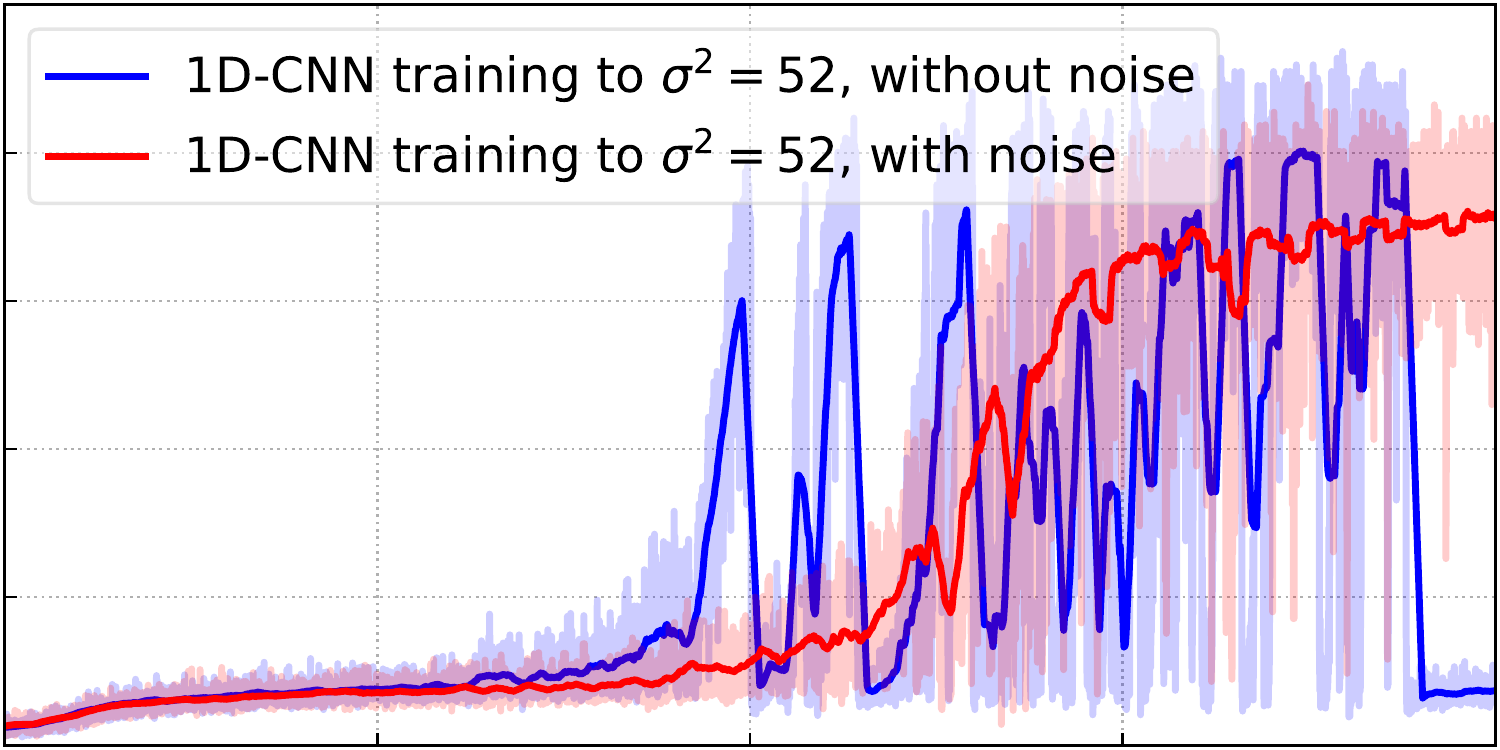}
  \caption{Learning curves of the 1D-CNN agent targeting Gaussian MWD with $\sigma^2 = 52$, trained on environments with and without simulated noises. Convention is as in \fig~\ref{fig:learning_fcnn}, with the horizontal axis having a range of 851896 episodes.}
  \label{fig:learning_gv52_nonoise_noise}
\end{figure}

Another interesting finding is that performance collapses during the training process may be alleviated by introducing noise to the training environment.  \fig~\ref{fig:learning_gv52_nonoise_noise} compares the learning curve of the 1D-CNN agent on the non-noisy environment with that on the noisy environment, both targeting a Gaussian MWD with $\sigma^2 = 52$.  Although, on the non-noisy environment, the agent can learn a high-reward strategy more quickly and achieve a slightly higher peak performance, the learning curve on the noisy environment is much steadier.  Intuitively, exposing the agent to noisy states increases its tolerance to abrupt changes in concentrations of observables and so may improve the generalization of the learned network.\cite{jim1995effects}  Moreover, introducing noise may also be regarded as a stochastic regularization technique.\cite{srivastava2014dropout,wager2013dropout}  Overall, introducing certain types of noise on the states and actions seems to have little adverse effect on the training while helping the agents achieve better generalization.

\subsection{Targeting MWDs with diverse shapes}
Beyond Gaussian MWDs, we also trained the 1D-CNN agent against a series of diverse MWD shapes.  We have chosen bimodal distributions as a challenging MWD to achieve in a single batch process.  Such bimodal distributions have been previously studied as a means to controlling the microstructure of a polymeric material.\cite{yan2015matrix,zheng2017useful,sarbu2004polystyrene}

\begin{figure}[ht]
\centering
\includegraphics[width=0.95\columnwidth]{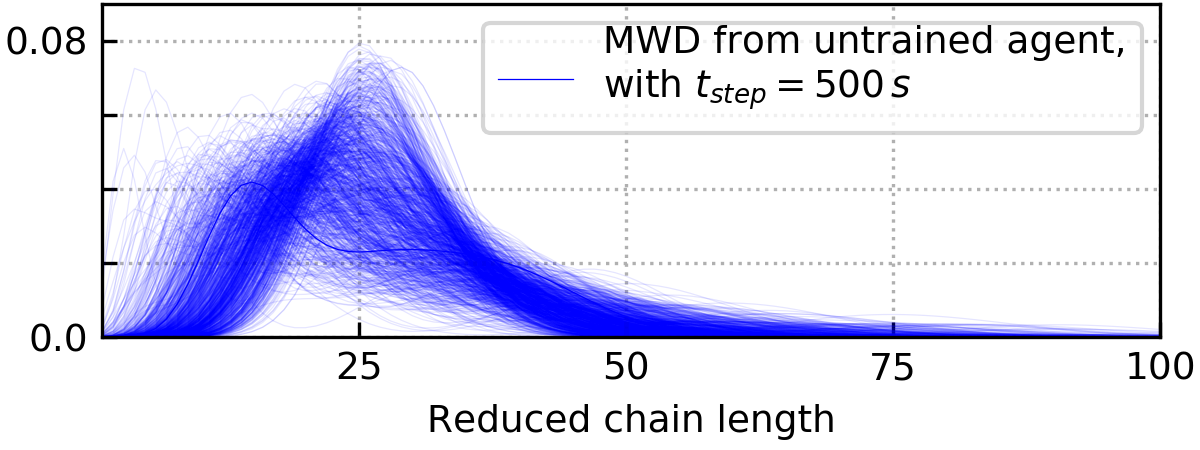}
\caption{Superposition of 1000 ending MWDs from untrained agents when the time interval between actions is 500 seconds. Vertical axis is fraction of polymer chains.}
\label{fig:is_untrained}
\end{figure}

To enable automatic discovery of control policies that lead to diverse MWD shapes, it is necessary to enlarge the search space of the RL agent, which is related to the variability in the ending MWDs generated by an untrained agent.  We found empirically that a larger time interval between actions leads to wider variation in the MWDs obtained with an untrained agent.  Throughout this section, the time interval between actions $t_\textit{step}$ is set to 500 seconds.  \fig~\ref{fig:is_untrained} shows 1000 superimposed ending MWDs given by the untrained agent with this new time interval setting, and the span is much greater than in \fig~\ref{fig:gv_untrained} where $t_\textit{step} = 100$ seconds.

\begin{figure}[b!]
  \centering
  \includegraphics[width=\effwidth]{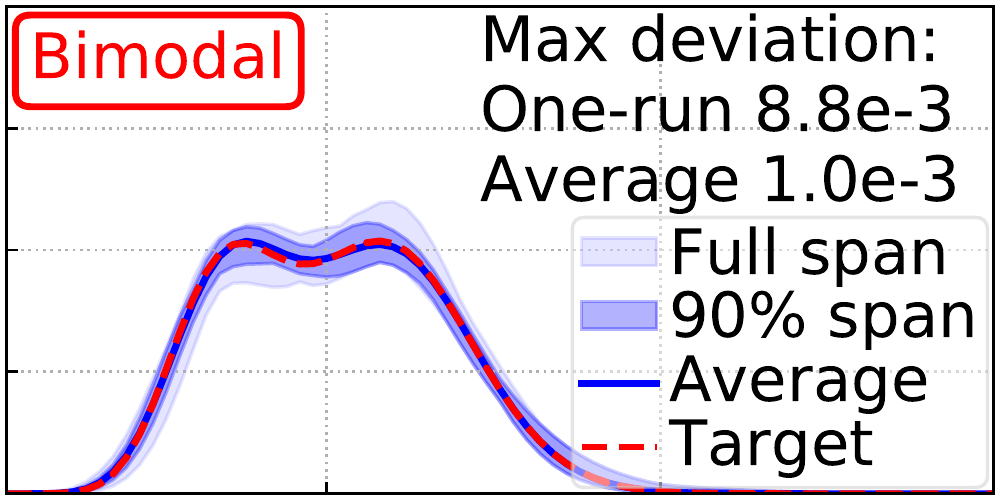}
  \includegraphics[width=\effwidth]{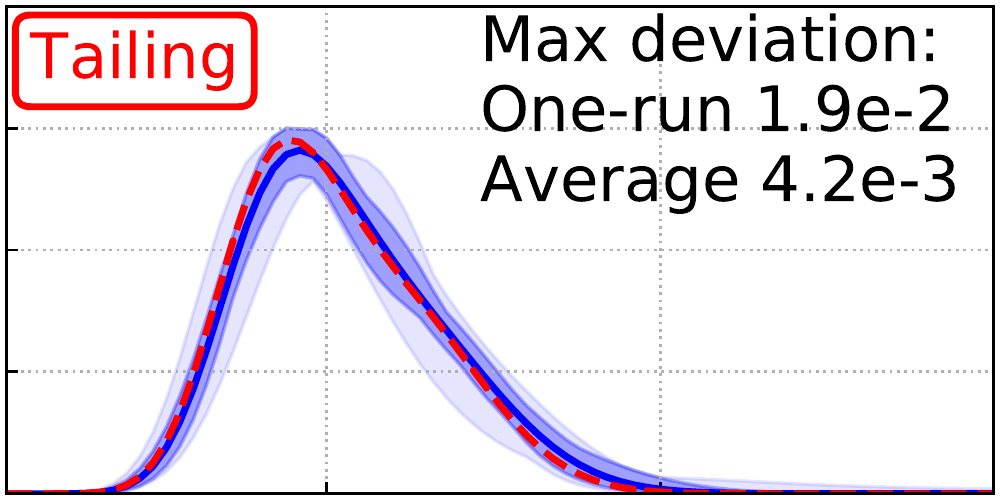} \\
  \includegraphics[width=\effwidth]{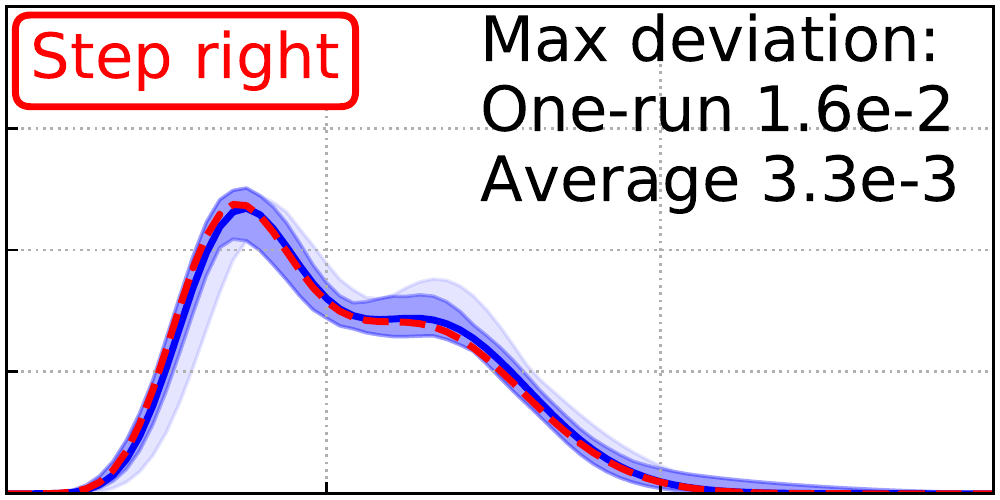}
  \includegraphics[width=\effwidth]{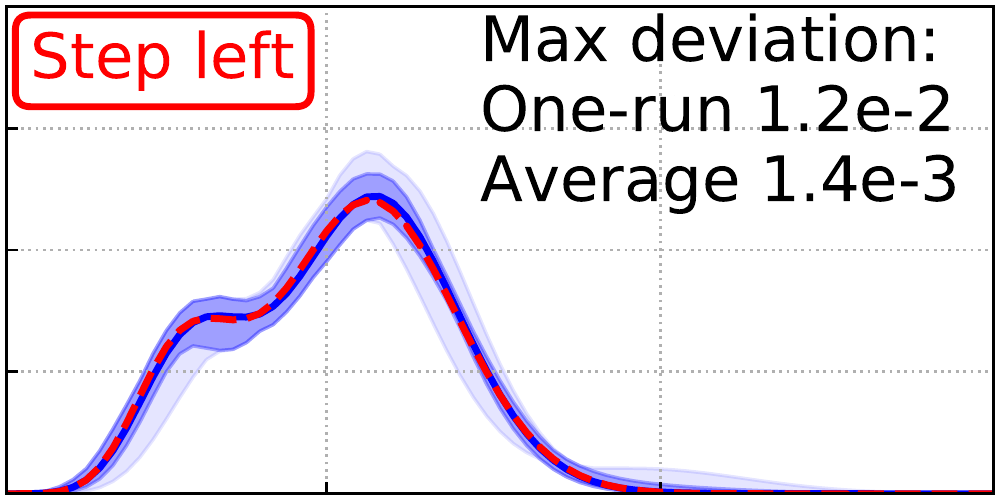} \\
  \includegraphics[width=\effwidth]{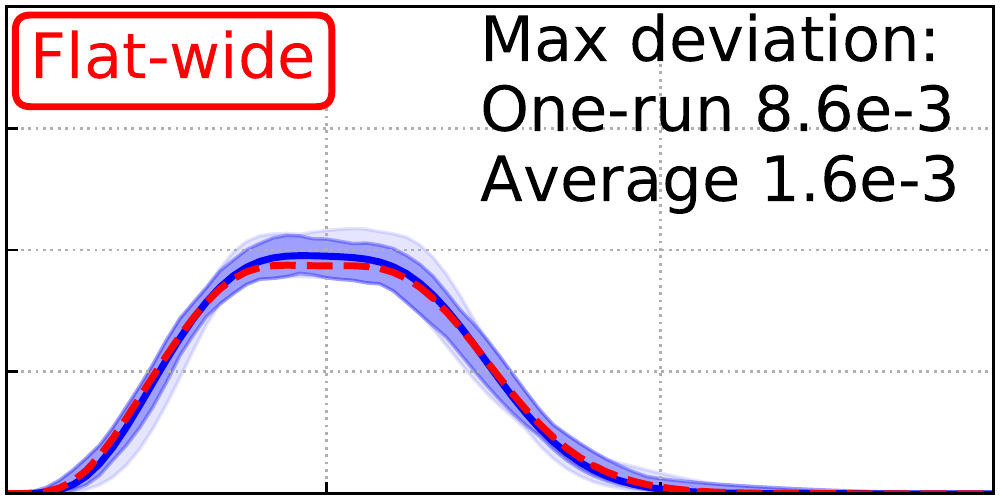}
  \includegraphics[width=\effwidth]{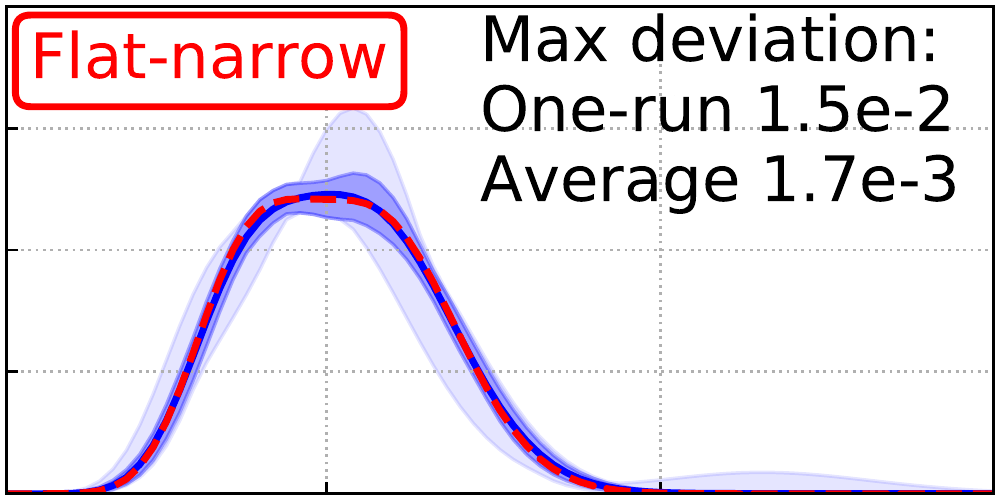}
  \caption{Performance of trained 1D-CNN agents on noisy, with-termination environments targeting diverse MWD shapes. In each subplot, the horizontal axis represents the reduced chain length and runs from 1 to 75, and the vertical axis is fraction of polymer chains and runs from 0.0 to 0.08.}
  \label{fig:is_effect}
\end{figure}

\begin{figure}[b!]
  \centering
  \includegraphics[width=\effwidth]{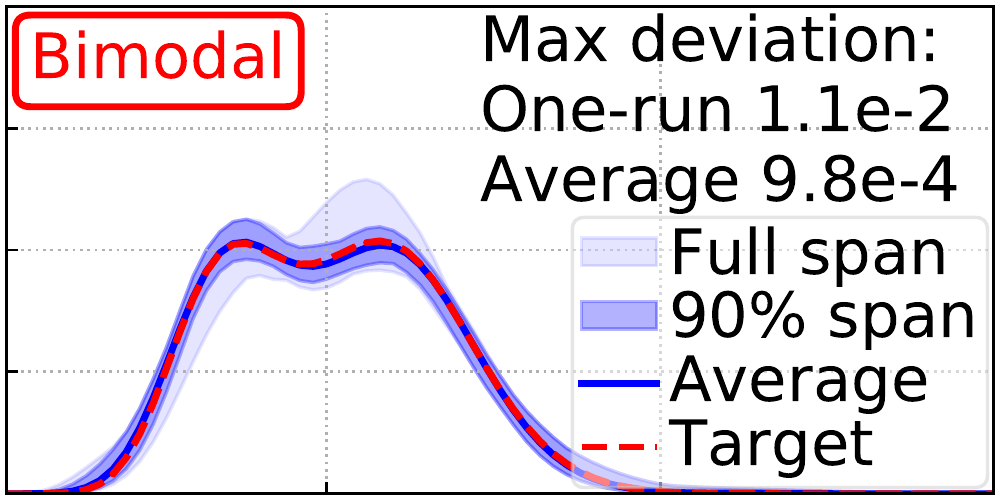}
  \includegraphics[width=\effwidth]{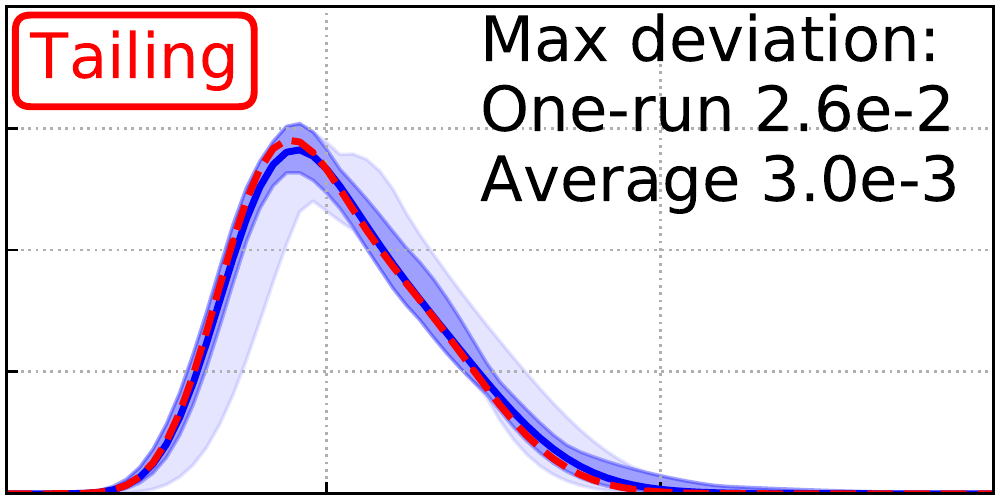} \\
  \includegraphics[width=\effwidth]{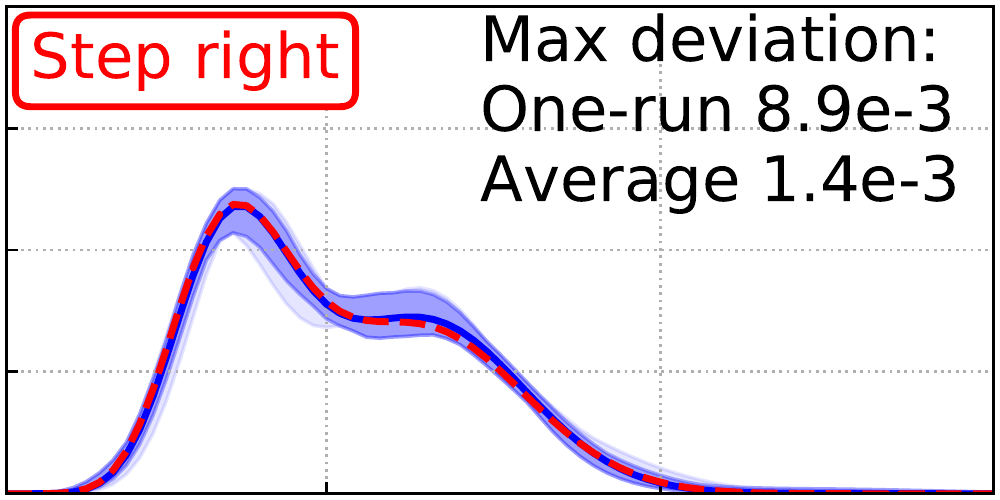}
  \includegraphics[width=\effwidth]{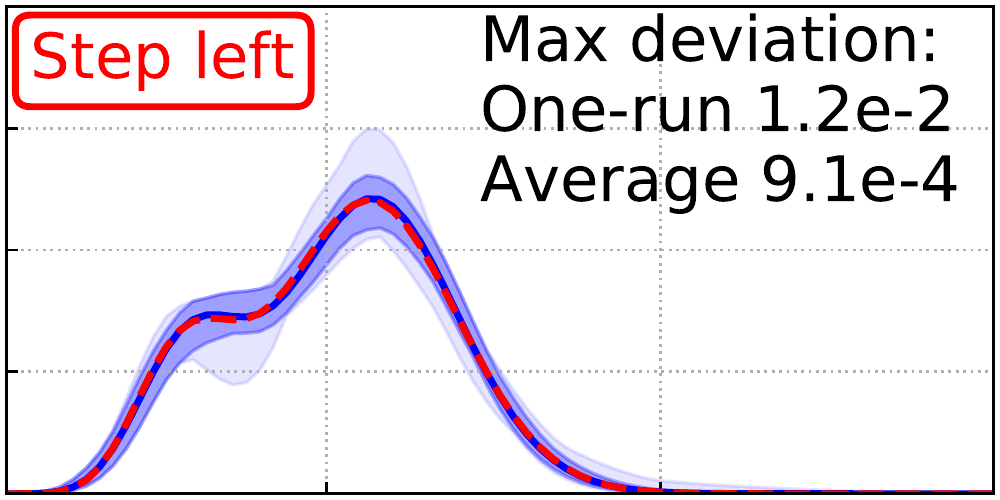} \\
  \includegraphics[width=\effwidth]{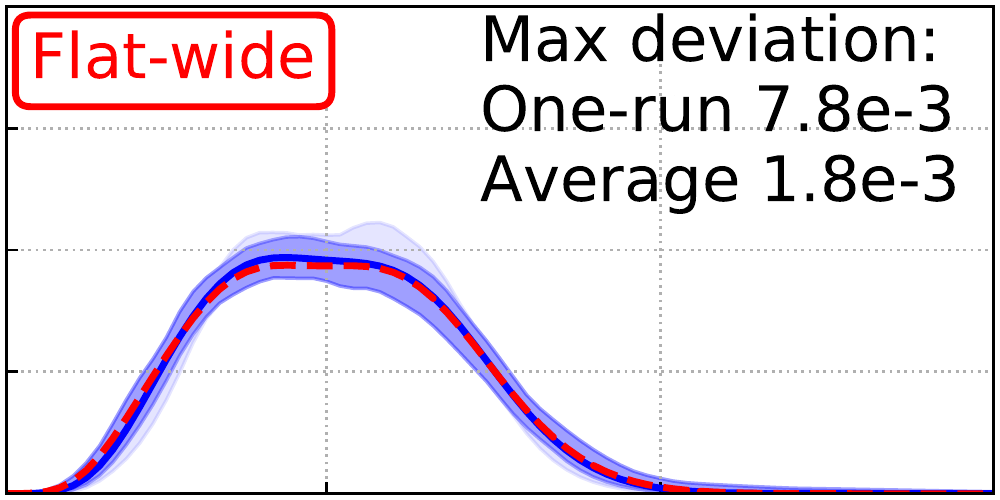}
  \includegraphics[width=\effwidth]{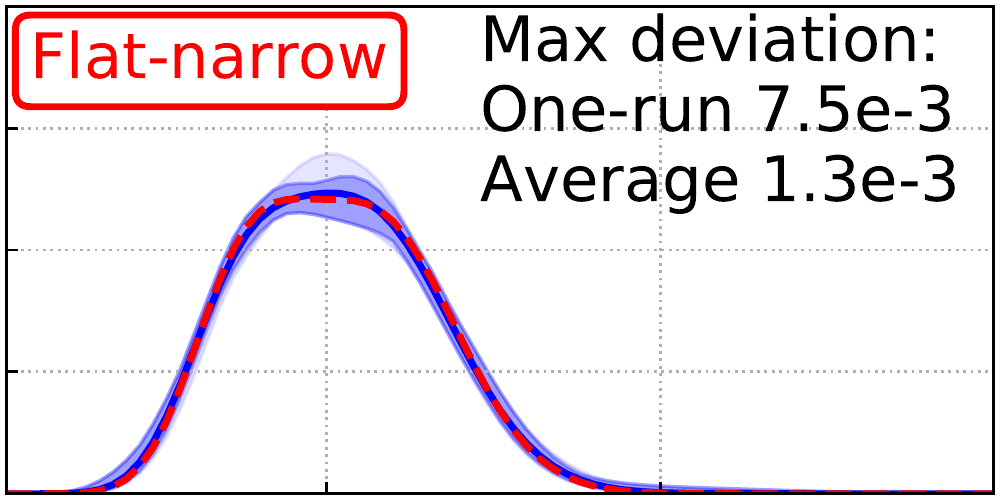}
  \caption{Performance of trained 1D-CNN agents on noisy, with-termination environments targeting diverse MWD shapes, where the chain propagation rate constant is increased by 100\% relative to the environments on which the agents were trained. Convention is as in \fig~\ref{fig:is_effect}.}
  \label{fig:is_dk_effect}
\end{figure}

The target MWDs with diverse shapes are manually picked from 1000 random ATRP simulation runs (i.e., episodes under the control of an untrained agent).  Agents trained on these targets have satisfactory performance.  The average MWDs over 100 batch runs match the targets nearly perfectly.  In addition, there is a large probability ($90\%$) that a one-run ending MWD controlled by a trained agent falls into a thin band whose deviation from the target is less than $1\times10^{-2}$ (\fig~\ref{fig:is_effect}).  All these agents are trained on noisy, no-termination environments and evaluated on noisy, with-termination environments.  The parameters specifying the noise are identical to those used in the earlier sections.  The results indicate that a simple convolutional neural network with less than $10^4$ parameters can encode control policies that lead to complicated MWD shapes with surprisingly high accuracy.  Again, adding noise to the states, actions, and simulation parameters does not degrade the performance of the RL agents significantly.  This tolerance to noise may allow transfer of control policies, learned on simulated reactors, to actual reactors.

To further investigate the potential transferability of the state-dependent control policies, we also evaluate the agents trained above on environments where the propagation rate constant $k_p$ is increased by $100\%$.  The other rate constants ($k_a$, $k_d$, and $k_t$) were held fixed, such that we are varying the relative time scales of the two interacting chemistries, propagation versus activation/deactivation, in ATRP (\fig~\ref{fig:atrp_scheme}).  The increase in chain propagation alters, for example, the average number of monomers added to an active chain before it is converted back to a dormant chain.  In applying the agents, the time intervals $t_\textit{step}$ and $t_\textit{terminal}$ are reduced by $50\%$ so that the reactions have a similar monomer conversion rate before and after the change to $k_p$.  As shown by \fig~\ref{fig:is_dk_effect}, this significant change in the ATRP reaction kinetics only slightly downgrades the agents' performance, with the average ending MWD remaining close to the target.  The successful transfer of agents trained on one set of kinetic parameters to a simulation with a different set of kinetic parameters suggests that having the agents base their decisions on the current state of the reaction leads to control policies that can transfer between chemical systems.

\section{Conclusion}
This paper introduces a general methodology for using deep reinforcement learning techniques to control a chemical process in which the product evolves throughout the progress of the reaction.  A proof-of-concept for the utility of this approach is obtained by using the controller to guide growth of polymer chains in a simulation of ATRP.  ATRP was chosen because this reaction system allows detailed control of a complex reaction process.  The resulting controllers are tolerant to noise in the kinetic rate constants used in the simulation, noise in the states on which the controller bases its decisions, and noise in the actions taken by the controller.  This tolerance to noise may allow agents trained on simulations of the reaction to be transferred to the actual laboratory without extensive retraining, although evaluation of this aspect is left to future work.  This approach, of carrying out initial training of a controller on a simulation, has been successfully applied in other domains such as robotics and vision-based RL.\cite{levine2016learning,christiano2016transfer,rusu2016sim}  Additional work is also needed to better understand the extent to which the controller can achieve synthetic targets when decisions are based on less detailed information regarding the state of the reactor.  The ability of the approach to target multiple properties,\cite{sprague2003multiple,van2014multi} such as targeting MWD and viscosity simultaneously, or targeting more complex architectures, such as gradient or brush polymers, also remains to be explored.  Our efforts to optimize the reinforcement learning methodology is still ongoing, and we hope to apply similar approaches to guide other chemical reactions.

A developmental open-source implementation of our approach is freely available on GitHub (\url{https://github.com/spring01/reinforcement_learning_atrp}) under the GPL-v3 license.



\bibliography{references}
\bibliographystyle{arxiv}

\end{document}